\documentclass[11pt,twoside]{article}

\usepackage{caption}
\usepackage{subcaption}
\usepackage{graphicx}
\usepackage{asp2021}
\usepackage{textcomp}

\aspSuppressVolSlug
\resetcounters

\usepackage{amsmath} 
\newcommand{\angstrom}{\text{\normalfont\AA}}

\newcommand{\ha}{\textrm{H}\ensuremath{\alpha}}
\newcommand{\hb}{\textrm{H}\ensuremath{\beta}}
\newcommand{\oiii}{[\textrm{O}~\textsc{iii}]}
\newcommand{\nii}{[\textrm{N}~\textsc{ii}]}
\newcommand{\sii}{[\textrm{S}~\textsc{ii}]}
\newcommand{\feii}{[\textrm{Fe}~\textsc{ii}]}
\newcommand{\caii}{[\textrm{Ca}~\textsc{ii}]}
\newcommand{\oiiilam}{[\textrm{O}~\textsc{iii}]~\ensuremath{\lambda5007}}
\newcommand{\niilam}{[\textrm{N}~\textsc{ii}]~\ensuremath{\lambda6584}}
\newcommand{\niidoub}{[\textrm{N}~\textsc{ii}]~\ensuremath{\lambda6548,\lambda6584}}

\graphicspath{{./}{figs/}}

\begin{document}

\title{Reconstructing and Classifying SDSS DR16 Galaxy Spectra with Machine-Learning and Dimensionality Reduction Algorithms}
\author{Felix Pat,$^{1,2}$ St\'ephanie Juneau,$^2$ Vanessa B\"ohm,$^3$ Ragadeepika Pucha,$^1$ Alex G. Kim,$^3$ A. S. Bolton,$^2$ Cleo Lepart,$^3$ Dylan Green,$^4$ and Adam D.\ Myers$^5$
\affil{$^1$University of Arizona, Tucson, AZ, USA; \email{felixpat10@arizona.edu,rpucha@arizona.edu}}
\affil{$^2$NSF's NOIRLab, Tucson, AZ, USA; \email{stephanie.juneau@noirlab.edu,adam.bolton@noirlab.edu}}
\affil{$^3$Berkeley Center for Cosmological Physics, University of California, Berkeley, CA 94720, USA; \email{vboehm@berkeley.edu,agkim@lbl.gov,clepart@lbl.gov}}
\affil{$^4$Department of Physics and Astronomy, University of California, Irvine, 92697, USA; \email{dylanag@uci.edu}}
\affil{$^5$Department of Physics and Astronomy, University of Wyoming, Laramie, WY 82071, USA; \email{amyers14@uwyo.edu}}
}


\paperauthor{Felix Pat}{felixpat10@arizona.edu}{}{University of Arizona}{Department of Physics}{Tucson}{AZ}{85719}{USA}
\paperauthor{St\'ephanie Juneau}{stephanie.juneau@noirlab.edu}{0000-0002-0000-2394}{NSF's NOIRLab}{}{Tucson}{AZ}{85719}{USA}
\paperauthor{Vanessa B\"ohm}{vboehm@berkeley.edu}{0000-0003-3801-1912}{Lawrence Berkeley National Laboratory}{Physics Division}{Berkeley}{CA}{94720}{USA}
\paperauthor{Ragadeepika Pucha}{rpucha@arizona.edu}{0000-0002-4940-3009}{University of Arizona}{Steward Observatory}{Tucson}{AZ}{85721}{USA}
\paperauthor{A.~G.~Kim}{agkim@lbl.gov}{0000-0001-6315-8743}{Lawrence Berkeley National Laboratory}{Physics Division}{Berkeley}{CA}{94720}{USA}
\paperauthor{A.~S.~Bolton}{adam.bolton@noirlab.edu}{0000-0002-9836-603X}{NSF's NOIRLab}{}{Tucson}{AZ}{85719}{USA}
\paperauthor{Cleo Lepart}{clepart@lbl.gov}{0000-0002-8740-4950}{Lawrence Berkeley National Laboratories}{Physics Division}{Berkeley}{CA}{94720}{USA}
\paperauthor{D.~Green}{dylanag@uci.edu}{0000-0002-0676-3661}{University of California, Irvine}{Department of Physics and Astronomy}{Irvine}{CA}{92697}{USA}
\paperauthor{Adam D. Myers}{amyers14@uwyo.edu}{}{Department of Physics and Astronomy}{University of Wyoming}{Laramie}{WY}{82071}{USA}

\begin{abstract}
Optical spectra of galaxies and quasars from large cosmological surveys are used to measure redshifts and infer distances. They are also rich with information on the intrinsic properties of these astronomical objects. However, their physical interpretation can be challenging due to the substantial number of degrees of freedom, various sources of noise, and degeneracies between physical parameters that cause similar spectral characteristics. To gain deeper insights into these degeneracies, we apply two unsupervised machine learning frameworks to a sample from the Sloan Digital Sky Survey data release 16 (SDSS DR16). The first framework is a Probabilistic Auto-Encoder (PAE), a two-stage deep learning framework consisting of a data compression stage from 1000 elements to 10 parameters and a density estimation stage. 
The second framework is a Uniform Manifold Approximation and Projection (UMAP), which we apply to both the uncompressed and compressed data. Exploring across regions on the compressed data UMAP, we construct sequences of stacked spectra which show a gradual transition from star-forming galaxies with narrow emission lines and blue spectra to passive galaxies with absorption lines and red spectra. Focusing on galaxies with broad emission lines produced by quasars, we find a sequence with varying levels of obscuration caused by cosmic dust. The experiments we present here inform future applications of neural networks and dimensionality reduction algorithms for large astronomical spectroscopic surveys.
\end{abstract}

\section{Introduction}\label{sec:intro}

Traditionally, researchers examined astronomical spectroscopic data by hand, and for large data sets, astronomers developed extensive automated spectroscopic analysis \citep[e.g.,][]{Almeida+2010}. One example of a large spectroscopic survey is the Sloan Digital Sky Survey \citep[SDSS;][]{York+2000}, which began its operations in the year 2000 and produced over 3.5 million unique spectra. Among the many past and ongoing efforts to analyze SDSS spectra, the use of machine learning is becoming increasingly popular. Machine learning is a powerful method for solving classification and regression problems, and the use of neural networks allows for complex data sets to be analyzed deeply. The last few years have seen a boom in their application to galaxy surveys \citep[see review by][and references therein]{Huertas-Company+2022}.

Astronomical spectroscopic lines are full of information including chemical abundance, stellar ionization, Active Galactic Nuclei (AGN) ionization, redshift and more. Stellar ionization may be produced by massive young stars, and AGN ionization may be produced from extremely luminous accretion disks around supermassive black holes with the most luminous being known as quasars. Spectra produced by the presence of accretion disks are characterized by broad emission lines, but a difficulty in identifying broad-lines occurs when the accretion disk is partly or fully obscured by surrounding dust. Therefore, we expect different astronomical objects to produce different spectral line features \citep{Osterbrock+2006}. 

A common way to determine the classification of objects is through a Baldwin, Phillips, Terlevich (BPT) diagnostic diagram \citep{Baldwin+1981}, which employs the following strong emission lines: \hb, \oiiilam\ (hereafter \oiii), \ha\ and \niilam\ (hereafter \nii). Specifically, the BPT diagram plots \(\log({\nii}/{\ha})\) versus \(\log(\oiii/\hb)\) to assess the dominant source of ionization. A commonly used classification scheme splits galaxies into star-forming, AGN, and composites based on their location with respect to demarcation lines proposed by \citet{Kewley+2001} and \citet{Kauffmann+2003}. 

The WHAN\footnote{Named after $\mathrm{W_{\ha}}$ versus \nii/\ha\ where $\mathrm{W_{\ha}}$ is the \ha\ equivalent width.} diagnostic diagram \citep{CidFernandes+2011} was proposed to further distinguish strong AGNs, weak AGNs, retired galaxies, and passive galaxies. \citet{CidFernandes+2011} argue that retired galaxies can present signatures of low ionization nuclear emission region (LINER) but are instead dominated by old stellar populations. Knowledge of the physical conditions within galaxies and the sources of ionization is crucial to our understanding of galaxy and black hole evolution. However, these emission line diagnostic diagrams focus on small regions in the entire spectrum. The hope of using machine learning is to utilize the whole spectrum for a more comprehensive analysis. 

Applications of deep learning, a subset of machine learning, on galaxy surveys include computer vision, inferring physical properties of galaxies, discovery, and cosmological simulations. Auto encoders (AEs) are well suited for data exploration, classification and discovery including tasks such as identifying outliers, reconstructing noiseless data, or visually inspecting complex datasets by projecting them to lower dimensions \citep{Huertas-Company+2022}. An AE is a two-part neural network consisting of an encoder and decoder. The encoder learns how to efficiently compress the input and it outputs the compressed data. The decoder aims to reconstruct the input through the compressed data, and together they optimize the compressed data by learning how to include the most crucial information for reconstruction in a specified limited space.

Previous work on spectra classification and regression by Variational Auto-Encoder \citep[VAE,][]{Portillo+2020} used their compressed data to construct sequences of synthetic spectra along a few parameters to show tracks of star-forming, extreme line emitting, post-starburst, and active galaxies. The VAE compressed data parameters are arbitrary, but some interpretations correlate with continuum slope, stellar age, emission and absorption line strength, and 4000\angstrom\ and Balmer break strength. Their VAE approach showed improvement to other dimensionality-reduction methods like Principal Component Analysis (PCA). However, some complications from the VAE include difficulty in exploring and interpreting the six-parameter compressed data, significantly underpredicting the uncertainty in the compressed data for low signal-to-noise (S/N) spectra and encoding bad pixels in the compressed data. 

Here, we use a Probabilistic Auto-Encoder \citep[PAE;][]{Boehm+2022}. Different to VAEs, this framework is trained in two-stages. In the first stage, an AE is trained to compress the SDSS spectra to ten parameters. In the second stage, a density estimator is trained to learn the probability distribution of the compressed data, which is useful for identifying common versus rare spectra such as outliers or anomalies (B\"ohm et al., in preparation). Together, the AE and the normalizing flow density estimator function as an efficient dimensionality reduction algorithm with a robust assessment of the probability distribution for the data set.

Another powerful dimensionality reduction algorithm is Uniform Manifold Approximation and Projection (UMAP), which acts as a general non-linear dimension reduction technique \citep{McInnes+2018}. Previous work used UMAP for classifying SDSS sources without spectra \citep{Clarke+2020}. In contrast, we specifically use UMAP on spectra. In the first case, we reduce the 1000-element uncompressed spectra to two dimensions. In the second case, we reduce the 10D AE compressed data to two dimensions. The motivations are to employ the full information from the spectra and to assess the topological impacts of the AE compression step. By comparing both UMAP projections with classification labels from traditional emission line diagnostics, we aim to determine which projection is best suited to identify patterns that translate to physical properties of galaxies and quasars.

In brief, we view the compressed SDSS DR16 data set produced by AE in different dimensions: the initial 10D from the AE as a corner plot, the 1D probability distribution from the density estimator as a histogram, and the 2D projection from UMAP as a scatter plot. From these three variations, we compare to a classification scheme from the BPT and WHAN diagrams as an alternative or complement to these standard emission line diagnostic diagrams. We identify interesting trends in the UMAP embedded space that could lead to follow-up studies, e.g., the transition from star-forming to passive galaxies or the variation among sub-populations (or phases) of quasars.

\section{SDSS DR16 Galaxy Sample}\label{sec:sdss}

The Sloan Digital Sky Survey data release 16 \citep[SDSS DR16,][]{Ahumada+2020} is a wide area spectroscopic survey with 3,518,265 unique spectra. Similar \citet{Portillo+2020}, we select a sample of galaxies and quasars at low redshift (0.05 to 0.36). We require a high-quality redshift with no warnings (`ZWARNING'=0). The selected spectra are preprocessed by shifting them to the rest frame. After applying the redshift correction, we resample the spectra on a \(\mathrm{log_{10}}\) scale to a fixed rest-frame wavelength grid of 1000 elements representing spectral pixel from 3388\angstrom\ to 8318\angstrom. Hence, there are 1000 bins for the flux, inverse variance, and binary mask values. The mask is set to one for valid pixels and zero for invalid (masked) pixels, which are defined based on the SDSS `AND\_MASK' being set or the inverse variance being zero or a region of the spectral grid not being covered once the spectrum is shifted to the rest-frame. We choose to grid our spectral range to 1000 elements for a reasonable data set size and increasing the number of elements may provide more information on emission line profiles at the cost of more computing resources. This resolution corresponds to 5.9\AA\ over the \ha\ and \niidoub\ region, which is sufficient to avoid blending between those lines (separated by $\sim$20\AA). Afterwards, we compute a global signal-to-noise (S/N) ratio over the full spectral range and apply a minimum cut of $\mathrm{S/N} > 50$ to ensure high quality spectra. Those selection and data quality cuts result in a total sample of 279,284 galaxies. We further divide the sample into 209,462 galaxies for training and 69,822 galaxies for validation, which is a 3:1 data set split. The resulting redshift distributions are displayed in Figure~\ref{fig:redshift_dist}. 

\articlefigure[scale=0.4]{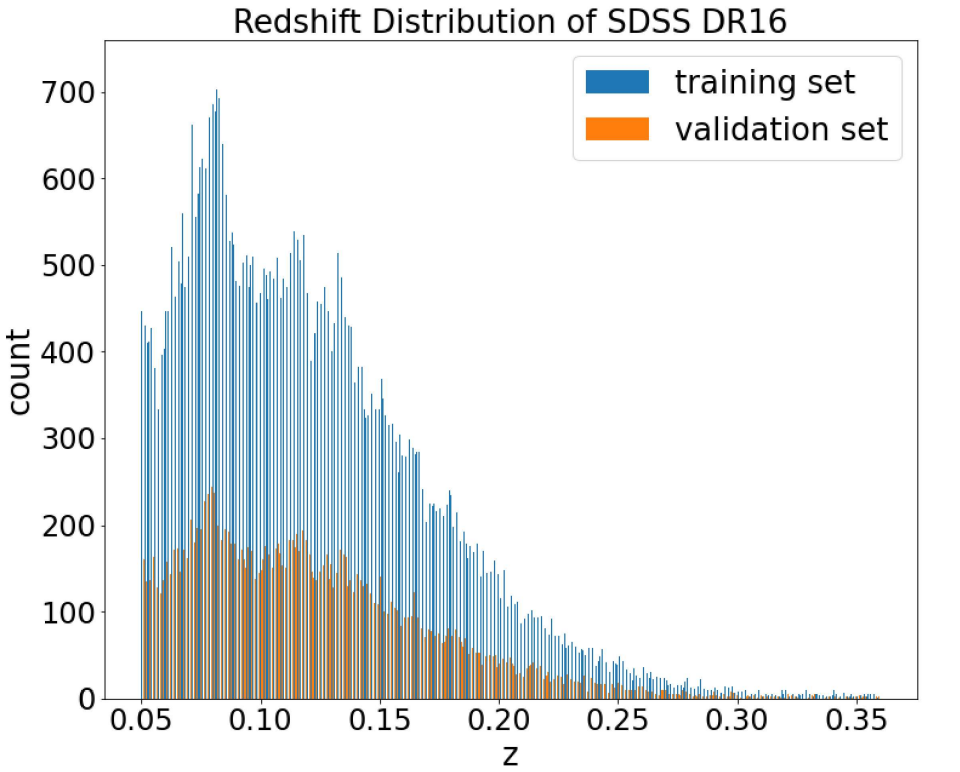}{fig:redshift_dist}
{Redshift distribution of SDSS DR16 galaxies from the training set (blue histogram) and validation set (orange histogram).}

Now, we have the flux ($10^{-17}{\mathrm{~erg\,s^{-1}\,cm^{-2}\,\angstrom^{-1}}}$), binary mask, flux noise (uncertainty), redshift, class (`STAR,' `GALAXY,' or `QSO'), subclass, signal-to-noise ratio, modified Julian date (MJD), plate number, and fiber identification number from SDSS DR16 \citep{Boehm+2022}. Lastly, we add a classification label based on the SDSS pipeline subclass, and on the BPT and WHAN optical emission line diagnostic diagrams (Section~\ref{sec:labels}). To obtain the necessary emission line measurements, we join our sample with the SDSS DR12 Portsmouth catalog \citep{Thomas+2013} through the plate, MJD and fiberid identifiers and retrieve the fluxes and associated uncertainties for the \hb, \oiii, \ha\ and \nii\ emission lines as well as the equivalent width (EW) for \ha. From the SDSS subclass, we only track whether the spectrum was tagged as having a `BROADLINE' according to the SDSS pipeline definition\footnote{“lines detected at the 10-sigma level with sigmas > 200 km/sec at the 5-sigma level" (Object Information section from \url{https://www.sdss.org/dr16/spectro/catalogs/})} \citep{Bolton+2012}. 

\section{Methodology}\label{sec:method}

\subsection{Labeling the Data}\label{sec:labels}

Using DR12 flux measurements for the \nii, \ha, \oiii, and \hb\ emission lines, we require a S/N$>$2 in at least three of the four lines for a given galaxy (with the fourth line having S/N$>$1) to assign a classification based on the BPT diagnostic diagram. Galaxies that fail these thresholds will be part of the passive galaxy category defined below. 
Plotting galaxies with emission lines onto the BPT diagnostic diagram shows three distinct regions labeled star-forming, AGN, and composite with demarcation lines determined by \citet{Kewley+2001} and \citet{Kauffmann+2003} (Figure~\ref{fig:bpt_whan}). Star-forming galaxies are defined as lying below the \citet{Kauffmann+2003} demarcation (solid line) and tend to have young stellar populations with narrow emission lines. AGN are active galaxies with either narrow lines (Type II) or sometimes both narrow and broad lines from active nuclei (Type I). They are defined as lying above the maximum starburst line reported by \citet[][dashed line]{Kewley+2001}. Composite galaxies are “between” star-forming galaxies and AGN with a significant likelihood to have mixed contributions from both processes. We adopt those three classes for galaxies that are not otherwise classified with a BROADLINE subclass according to the SDSS pipeline. The latter are instead labeled as broad-line galaxies. While we do not use the BPT diagram to further label the broad-line galaxies, we note that their narrow line ratios tend to occupy the AGN branch of the BPT (middle panel of Figure~\ref{fig:bpt_whan}).

\articlefigure[scale=0.345]{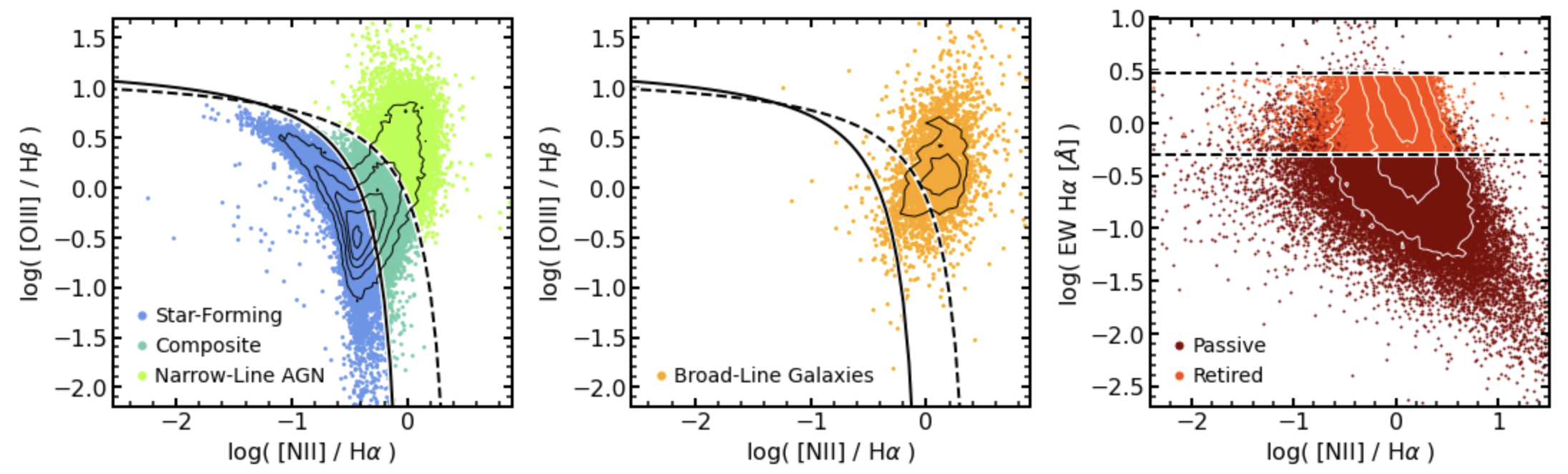}{fig:bpt_whan}{Traditional emission-line diagnostic diagrams applied to the combined training and validation sets. The BPT diagram is shown for the narrow-line spectra (left panel), and broad-line spectra (middle panel). The WHAN diagram is shown for retired and passive galaxies (right panel). The points are color-coded by the label as listed in the legends, and the contours indicate data point density with a logarithmic spacing. The small percentage of objects lacking a classification (labeled `N/A') are not shown. The BPT demarcation lines are from \citet[][solid line]{Kauffmann+2003} and \citet[][dashed line]{Kewley+2001} while the horizontal dashed lines on the WHAN diagram correspond to the \ha\ EW at 3\AA\ and 0.5\AA.}

However, since not all galaxies have high flux emission lines, we use other spectral features like the equivalent width of \ha\ to differentiate between passive and retired galaxies. \citet{CidFernandes+2011} defined retired galaxies as “galaxies that have stopped forming stars and are actually ionized by the hot low-mass evolved stars contained in them." While both retired and passive galaxies are dominated by old stellar populations, the former have weak emission lines and the latter do not have significant emission line detections (sometimes called “lineless" galaxies). We use \ha\ equivalent width (EW) cuts that are similar to the WHAN diagram \citep{CidFernandes+2011} with retired galaxies having \(0.5\angstrom < \ha\ \mathrm{EW} < 3\angstrom\) and passive galaxies having \(\ha\ \mathrm{EW} < 0.5\angstrom\). In our approach, retired galaxies were also required to satisfy $\mathrm{S/N}\ (\ha\ \mathrm{EW})>2$ or otherwise were labeled as passive. Their bivariate distributions on the WHAN diagram are shown in the right-hand panel of Figure~\ref{fig:bpt_whan} with horizontal lines corresponding to \ha\ EW at 3\AA\ and 0.5\AA.

Finally, a small percentage (0.14\%) of objects were labeled as `N/A' (not available) because they could not be classified due to being neither part of the DR12 emission line catalog nor having a BROADLINE subclass. In the end, our labels with the number of galaxies split between (training, validation) are N/A (306, 86), star-forming (63954, 21334), composite (26665, 8927), narrow-line AGN (11218, 3790), broad-line (6610, 2148), retired (39498, 13087), and passive (61211, 20450) galaxies. 

\subsection{PAE Architecture}\label{sec:nn_pae_arch}

Machine learning models learn based on the idea of repeating an experiment many times for improving precision, and each iteration through the entire training data set is called an epoch. The PAE is a combination of an auto-encoder and a neural density estimator, and three input variables are used for training: flux, binary mask, and flux uncertainty. Although we have the classification labels available, they are not used as part of training for any of the steps; the labels color-code clustering projections and identify spectra. 

With the auto-encoder, we integrate a Fully Connected Neural Network architecture defined by all nodes in one layer to connect to the nodes of the next layer. By connecting all the nodes, the weights of each layer during training can be fed backwards and forwards to increase learning efficiency and convergence speed. 

The density estimator adds a probabilistic structure to the auto-encoder. For the density estimation, we use a Sliced Iterative Normalizing Flow algorithm \citep[SINF;][]{Dai+2020}. SINF generative models are trained to find a bijective mapping between tractable distributions, such as a multivariate Gaussian, and the complex data distributions. Following the approach by \citet[][, and B\"ohm et al., in preparation]{Boehm+2022}, we use a Gaussianizing Iterative Slicing (GIS) SINF model to map the AE compressed data to a single probability density function.

Since our flux binning for SDSS is 1000 pixels, the input and output tensor for the AE corresponds to 1000 by the number of spectra for training or testing. The encoder and decoder each has two hidden layers with 800 and 590 nodes (Figure \ref{fig:nn_info}), and we use the Adam optimizer \citep{Kingma+2014} with a rectified linear activation function (ReLU) for forward propagation and back propagation.

Our choice of using a mask \(\chi^{2}\) loss function is motivated by accounting for measurement errors in the spectra (flux noise) and the reconstruction accuracy in the neural network’s training. To exclude the training of bad pixels, we incorporate the pixelwise binary mask into the loss function to avoid encoding bad pixels in the compressed data:

\begin{equation}\label{eq:loss_func}
{\chi_{loss}}^2=\frac{1}{1000}\sum_{\mathrm{pixel=0}}^{999} (\mathrm{F_{pixel}^{recon}-F_{pixel}^{data})^2\cdot{F_{pixel}^{noise}}\cdot{F_{pixel}^{mask}}}
\end{equation}
where $\mathrm{F_{pixel}^{recon}}$ is the reconstructed spectrum flux, $\mathrm{F_{pixel}^{data}}$ is the measured flux, $\mathrm{F_{pixel}^{noise}}$ is the measurement error, and $\mathrm{F_{pixel}^{mask}}$ is the binary mask. A perfect model fit converges toward a loss value of 1, and the loss function indicates when a model is overtraining if the loss value shows an increasing trend after reaching a global minimum. After hyperparameter fine-tuning, the training batch size is 32, valid batch size is 512, and learning rate is $10^{-3}$ with an exponential learning rate decay gamma factor of 0.99. The PAE is structured such that the AE and GIS can be optimized separately through different epochs and loss functions. We train the AE and GIS for 200 epochs each for convenience even though GIS has the option to stop automatically based on its own loss function trend.

\subsection{UMAP Architecture}\label{sec:umap}

UMAP is a two-step algorithm that creates a weighted k-neighbor graph and then projects a low dimensional layout of the graph \citep{McInnes+2018}. To use UMAP, the data must be uniformly distributed, locally connected, and locally constant by the Riemannian metric \citep{McInnes+2018}, which we assume SDSS spectra satisfy.
UMAP creates a low dimensional projection by transforming the k-neighbor graph to be topologically closed based on its local distance defined by the Riemannian metric. Then, UMAP optimizes a cross-entropy function fit and conveniently preserves global structure. The output is a Riemannian manifold projection with arbitrary axes units ranging from two dimensions to one-hundred dimensions. The advantages of UMAP compared to other dimensionality reduction algorithms like t-distributed stochastic neighbor embedding (t-SNE) are faster run time performance for larger data sets, flexible output dimension options, and preserving more global structure. 

While UMAP makes hyperparameter fine-tuning flexible, we start with the default options for the size of each local neighborhood equal to 15, effective minimum distance between points equal to 0.1, and metric as Euclidean space \citep{McInnes+2018}. UMAP, being a general dimensionality reduction algorithm, is applicable on the uncompressed and compressed data, and we qualitatively explore how changing the UMAP input changes the projection shape and separation of classes.

\section{Stacking SDSS Spectra}\label{sec:stack}

From UMAP, we apply criteria based on patterns identified by visually inspecting the classification labels, local data point density, and overall shape. In one study, we build sequences from passive to star-forming galaxies (Section~\ref{sec:results_sf_passive}), and in the other study, we focus on the BROADLINE subclass (Section~\ref{sec:results_bl}). We retrieve galaxies in each selected region of interest and return the galaxies' indices, plate numbers, MJDs, and fiber ID numbers. Using the Astro Data Lab\footnote{https://datalab.noirlab.edu} \citep{Fitzpatrick+2014,Nikutta+2020}, we crossmatch this information with the SDSS DR16 specObj table to retrieve the galaxies' spectrum identification numbers (specObjID) and redshifts for spectral stacking in the rest frame.

The original SDSS spectra have a fixed spacing of 0.0001 dex in the $\mathrm{log_{10}}$ wavelength space, which is convenient to compute the flux-conserved stacked spectra. We consider a reference log wavelength array, $\log~\lambda$, ranging from 3.5 to 3.9, in 0.0001 dex bins. For each galaxy in a given region of interest, we convert the accessed observed spectra (wavelength, flux, and inverse variance) to the rest-frame spectra using the spectroscopic redshift. Even with this transformation, the spacing of these spectra in the log wavelength scale remains constant. This allows us to align the $\log~\lambda$ of the spectra with the reference $\log~\lambda$ array without interpolation. When applicable, we pad the empty regions on either end of the spectra with missing values (Not a Number) for the flux and with zeros for the inverse variance. Finally, we calculate the stacked spectra of the galaxies to be the weighted average of the flux arrays, weighted by the inverse variance. This procedure was adapted from a publicly available example Jupyter Notebook that illustrates SDSS galaxy stacking\footnote{https://github.com/astro-datalab/notebooks-latest/blob/master/03\textunderscore ScienceExamples/EmLineGalaxies/ 01\textunderscore EmLineGalaxies\textunderscore SpectraStack.ipynb}.

\section{Results}\label{sec:results}
\subsection{Auto-Encoder's Impact on UMAP Projection}\label{}

We present two UMAP projections obtained from the same training set. In the first case, we applied the algorithm directly to the uncompressed spectra (with 1000 elements) while in the second case, we applied it to the AE compressed spectra (10D). In both cases, the label information is not considered in the training and is strictly used to color-code the points in the resulting 2D projection (Figure~\ref{fig:compare_umap}). 

\articlefiguretwo{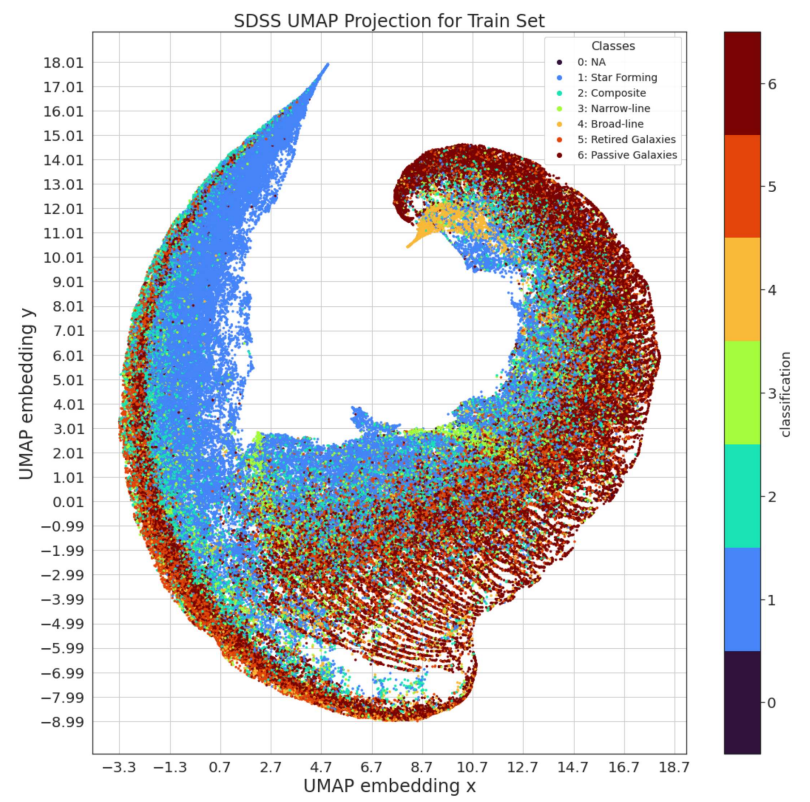}{figs/10D_input_umap}{fig:compare_umap}{UMAP projections of the SDSS training set in 2D Euclidean space.  \emph{Left:} UMAP with the uncompressed SDSS data.  \emph{Right:} UMAP with the 10D AE compressed data. The axes are expressed in arbitrary units. Each dot is color-coded to the classification label as defined by the legend and color bar.}\label{compare_umap}

Passing the uncompressed SDSS data into UMAP, we see in the left panel of Figure \ref{fig:compare_umap} that the UMAP projection separates star-forming (blue) and passive galaxies (dark red) the most strongly. Retired galaxies (light red) occupy a different projected space than the bulk of passive galaxies. There are a few notable features in the uncompressed SDSS UMAP (left panel): passive galaxies are characterized by striations branching from star-forming points, narrow-line AGNs (lime green) occupy two distinct bands of points, and the broad-line class (orange) is distinct with a topology that includes a narrowly defined branch of points. 

The motivation to input the 10D AE compressed data into UMAP is to qualitatively improve the separation of classes and effectiveness because the AE also trains on the mask and noise. From the right panel of Figure \ref{fig:compare_umap}, we see the AE compressed data provides more clearly defined regions separated by class through a different topology, and UMAP retains global features like separating retired and passive galaxies. We also note outlier candidates that appear as islands of mixed classes. While we do not investigate them further in this work, we note that their number and shapes vary with the choice of UMAP hyperparameters. These islands could be the subject of future work to determine if they comprise rare, unusual objects or spectral features due to data artifacts. In the remainder of this work, we focus on the UMAP projection of the AE compressed data set because it is easier to interpret due to presenting more obvious groupings and gradients based on class distributions. 

\subsection{Trends Between Star-forming and Passive Galaxies}\label{sec:results_sf_passive}

We reproduce the 2D UMAP projection for the compressed data set in Figure~\ref{fig:umap_topbot} to define regions of interest for investigation. In addition to a clean separation of star-forming galaxies (blue points) from retired and passive galaxies (in light red and dark red points respectively), we note that the latter form two distinct groups of points well separated

\articlefigure[scale=0.5]{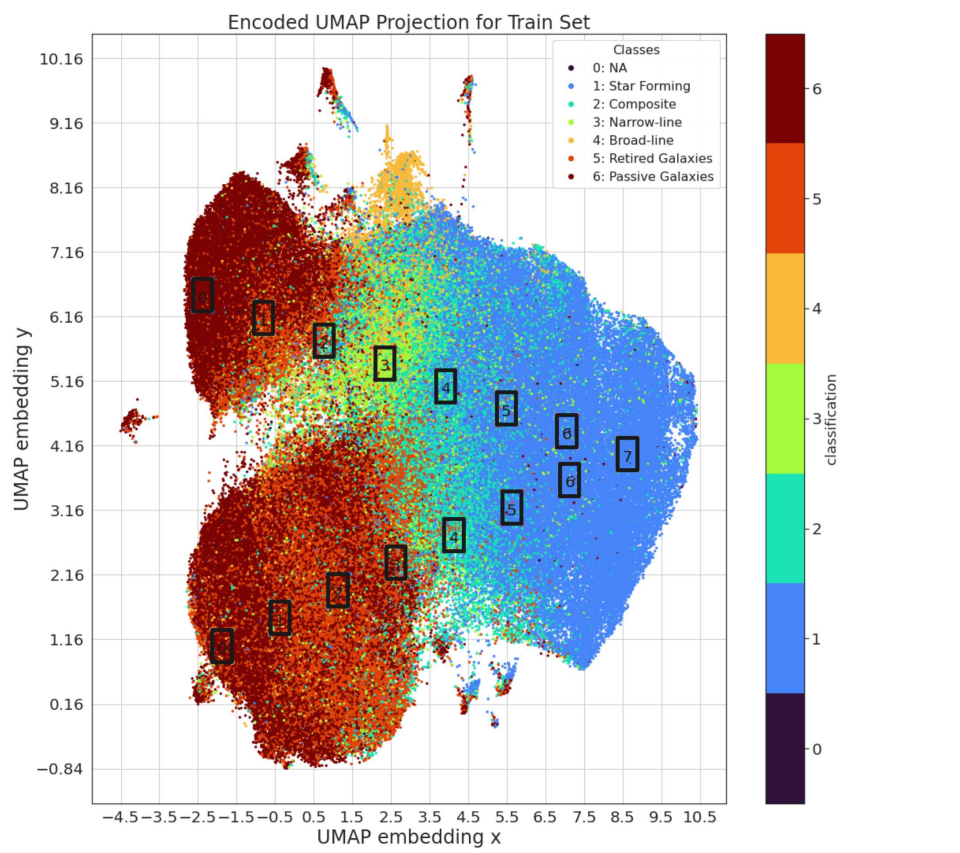}{fig:umap_topbot}
{Two-D Euclidean geometry UMAP projection after training on 10D compressed data reproduced from the right panel of Figure~\ref{fig:compare_umap}. The color bar represents the classification with the label names in the legend. The boxes select two sequences of stacked spectra of interest and are labeled from 0 to 7 from the left-hand side toward the right-hand side. Box 7 is in common to both sequences.}

\begin{figure}[!htb]
    \begin{subfigure}[t]{\textwidth}
    \centering
        \includegraphics[width=\linewidth]{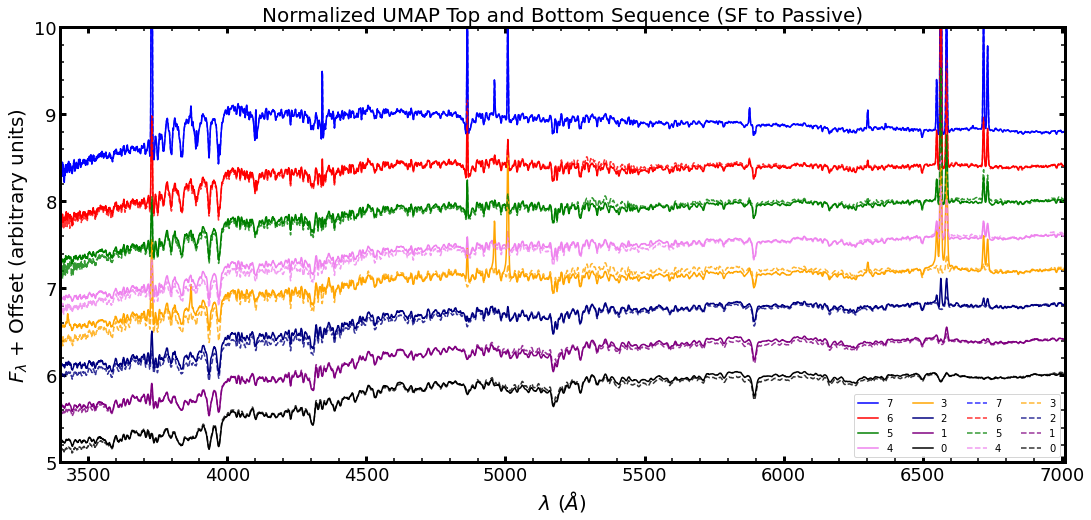} 
        \caption{Stacked spectra of top and bottom sequence from UMAP (Figure~\ref{fig:umap_topbot})} \label{fig:topbot_all}
    \end{subfigure}
    \vspace{0.25cm}
    
    \centering
    \begin{subfigure}[t]{0.48\textwidth}
        \centering
        \includegraphics[width=\linewidth]{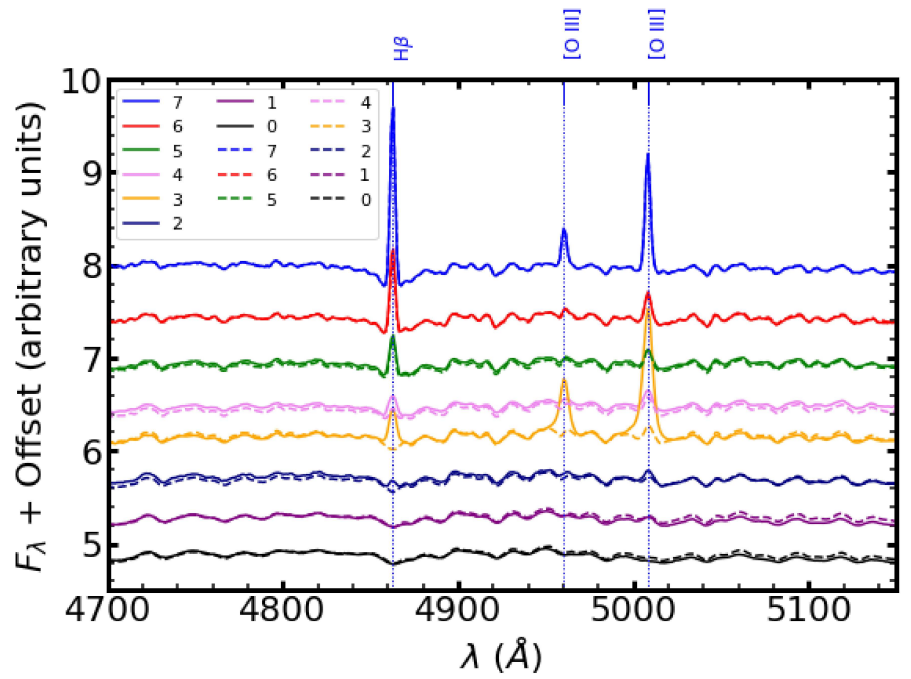} 
        \caption{Zoomed in stacked spectra around \hb\ and \oiii\ region} \label{fig:topbot_hb}
    \end{subfigure}
    \hfill
    \begin{subfigure}[t]{0.48\textwidth}
        \centering
        \includegraphics[width=\linewidth]{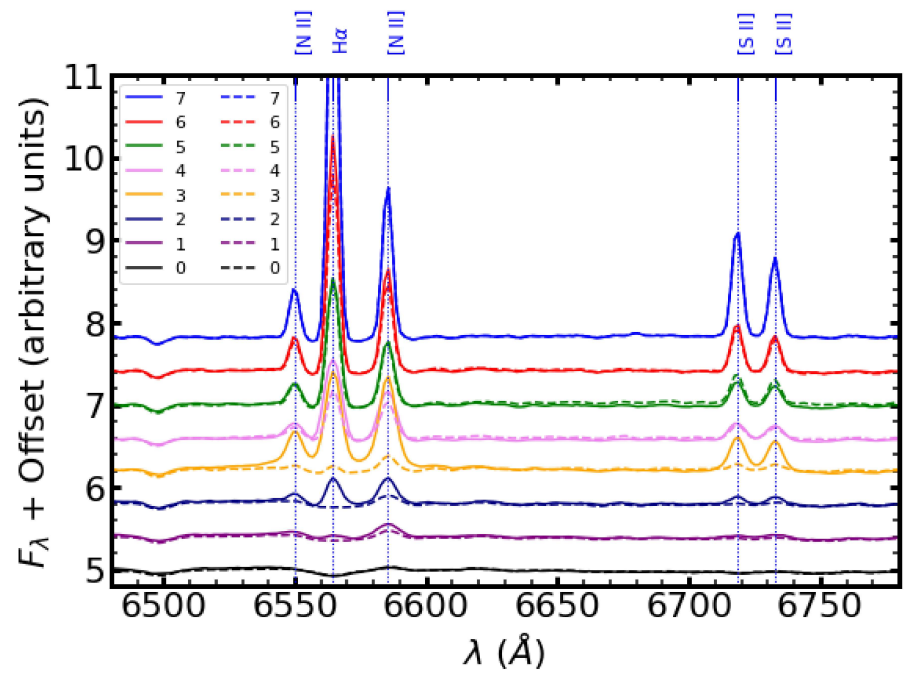} 
        \caption{Zoomed in stacked spectra around \ha\ and \nii\ region} \label{fig:topbot_ha}
    \end{subfigure}
    
    \caption{From the regions selected on UMAP (Figure~\ref{fig:umap_topbot}), the stacked spectra in Figure \ref{fig:topbot_all} are ordered from star-forming to retired and passive galaxies with solid lines being the top sequence and dashed lines being the bottom sequence. Note that spectrum \#7 is in common to both sequences. Figure \ref{fig:topbot_hb} zooms in around \hb\ and Figure \ref{fig:topbot_ha} zooms in around \ha. In panels (b) and (c), the location of typical emission lines is marked with vertical dotted lines and the transitions are labeled at the top.}\label{fig:topbot_spec}
\end{figure}

\noindent from each other. The top group transitions through narrow-line AGN (lime green points) and composite galaxies (green points) before connecting to the dominant star-forming population (blue points). 

We define a series of 8 boxes to form a sequence, which we will use to stack spectra within each box (Figure~\ref{fig:umap_topbot}). Similarly, the bottom group of passive (dark red) and retired (light red) galaxies appears to connect to the star-forming population via a transition region dominated by composite galaxies (green). However, we note differences relative to the top group including an apparent higher prominence of retired galaxies (light red) and a lack of active galaxies including both narrow-line AGN (lime green) and broad-line galaxies (orange). We define a second series of 8 boxes (labeled from 0 to 7) to form a bottom sequence that converges with the top sequence on the star-forming galaxies side (box 7). In short, the top and bottom sequences mostly differ in the y-direction of the UMAP projection, and the box numbers increase from left to right along the x-direction.

For each boxed region in the 2D UMAP projection, we construct a corresponding stacked spectrum following the procedure described in Section~\ref{sec:stack}. We show them sequentially in Figure \ref{fig:topbot_all} to reveal patterns along a given sequence as well as to compare the two sequences to each other. Note that the top sequence are solid lines, and the bottom sequence are dashed lines. The continuum shape can be characterized by a slope (color) and the presence (or absence) of breaks, which are influenced by stellar ages, stellar metallicity, and the presence of dust. The strengths and widths of emission lines encode information on the source of ionization, the gas-phase metallicity, and the kinematics of the ionized gas. Focusing on the emission lines used in the BPT diagnostic diagram (Figure \ref{fig:topbot_spec}b,c), we see clear spectral differences and patterns between the two extremes of the sequences defined by the star-forming and passive galaxies, but we also note that intermediate cases dominated by composite and narrow-line AGN stand out (stacked spectrum from box 3 in the top sequence, shown with the solid yellow line). To compare stacked spectra from UMAP to reconstructed spectra from GIS, refer to Appendix \ref{app:pae}.

\subsection{Trends from Broad-Line Galaxies}\label{sec:results_bl}

Spectra with broad-lines seem to be categorized in their own space (orange dots in  Figure~\ref{fig:umap_topbot}). They form a narrow branch protruding from their main locus, and spread

\begin{figure}[!htb]
    
    \begin{subfigure}[t]{\textwidth}
    \centering
        \includegraphics[width=\linewidth]{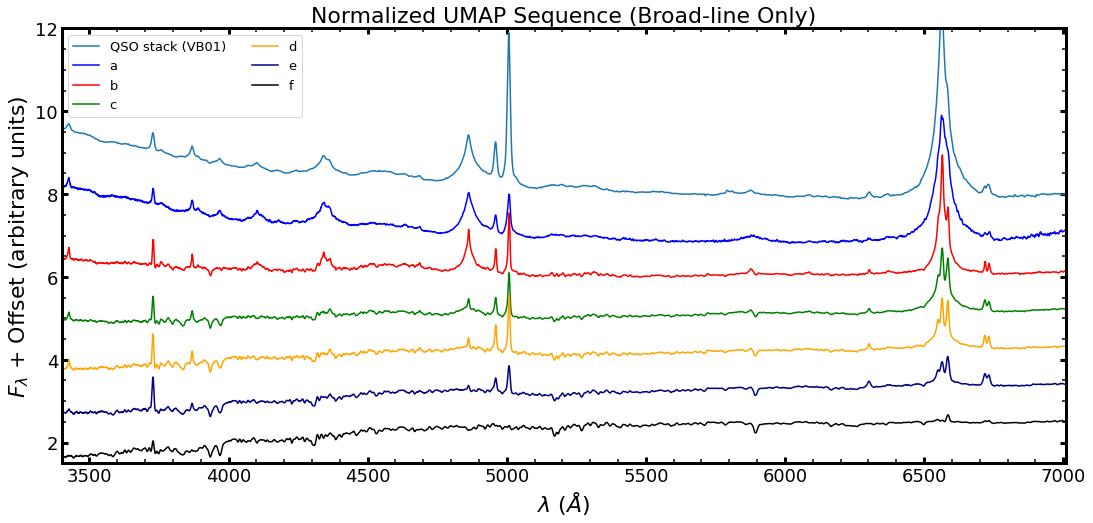} 
        \caption{Stacked spectra of broad-line sequence from UMAP (Figure~\ref{fig:umap_bl})} \label{fig:bl_all}
    \end{subfigure}
    \vspace{0.25cm}
    
    \centering
    \begin{subfigure}[t]{0.48\textwidth}
        \centering
        \includegraphics[width=\linewidth]{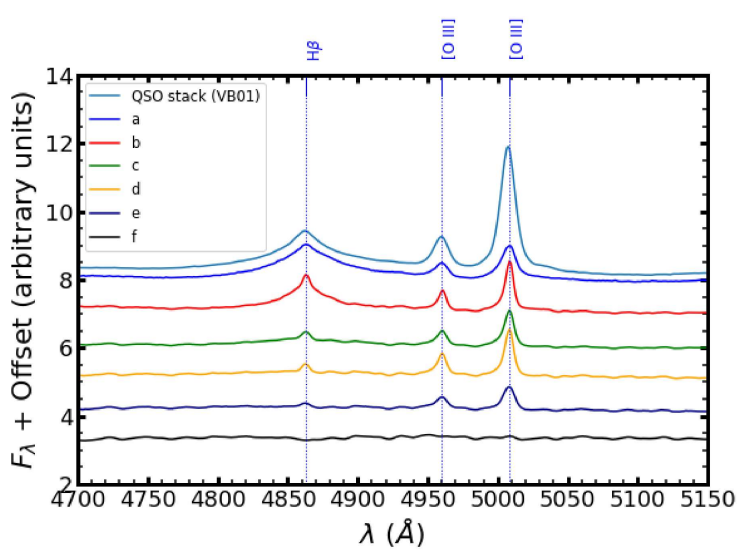} 
        \caption{Zoomed in broad-line stacked spectra around \hb\ and \oiii\ region} \label{fig:bl_hb}
    \end{subfigure}
    \hfill
    \begin{subfigure}[t]{0.48\textwidth}
        \centering
        \includegraphics[width=\linewidth]{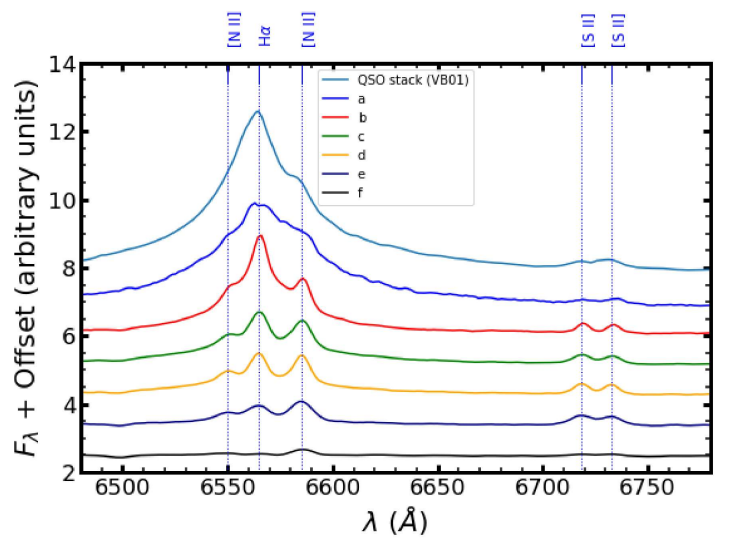} 
        \caption{Zoomed in broad-line stacked spectra around \ha\ and \nii\ region} \label{fig:bl_ha}
    \end{subfigure}
    
    \caption{From the regions selected on UMAP (Figure~\ref{fig:umap_bl}), the stacked spectra in Figure \ref{fig:bl_all} are ordered from the \citet{Berk+2001} reference (VB01) to regions `A' to `F' which are ordered from top to bottom on the UMAP projection (Figure \ref{fig:umap_bl}). Figure \ref{fig:bl_hb} zooms in around \hb\ and Figure \ref{fig:bl_ha} zooms in around \ha.}\label{fig:bl_spec}
\end{figure}

\articlefigure[scale=0.5]{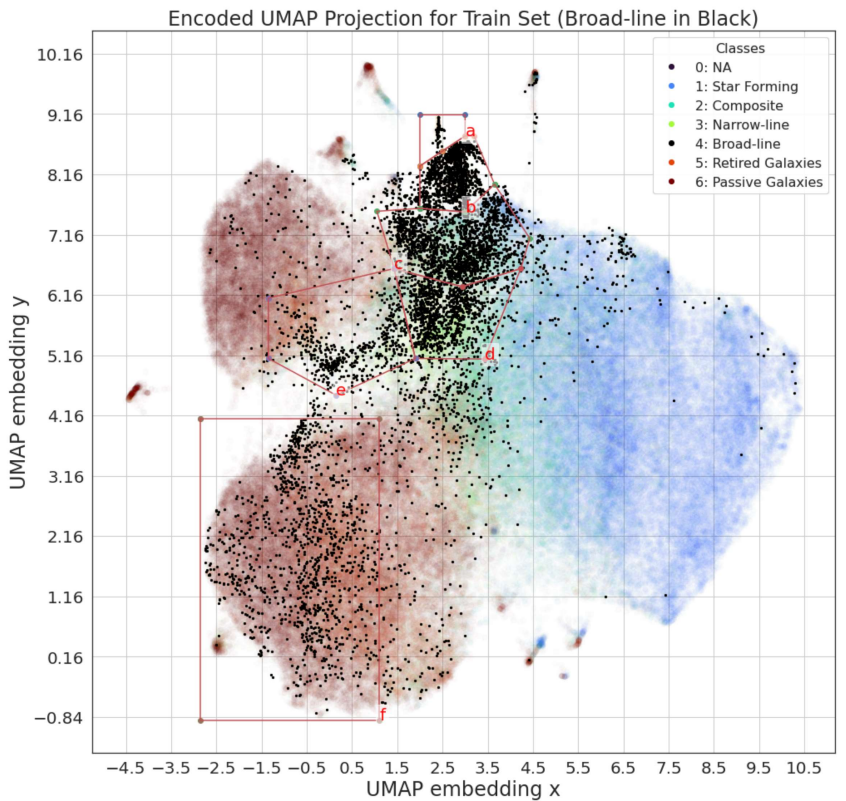}{fig:umap_bl}
{Two-D Euclidean geometry UMAP projection after training on 10D compressed data. The color bar represents the classification with the names in the legend, and broad-line galaxies have been changed to black to highlight that specific class. The polygons select the sequence of stacked spectra of interest in Figure~\ref{fig:bl_spec}.}

\noindent to mix with star-forming, composite, and narrow-line galaxies. Subsequent results will only include broad-line galaxies and their spectra with the same UMAP legend apart from broad-line galaxies being marked as black points in Figure~\ref{fig:umap_bl}. With the exception of a few of outliers, UMAP separates broad-line galaxies as a distinct classification from retired and passive galaxies. 

To learn about variations between subsets of galaxies labeled as having broad lines, we define areas based on the topology of the points on Figure~\ref{fig:umap_bl}. We visually select the regions by the most distinct narrow branch of broad-line points (`A'), the next nearest clustered group (`B'), and then defining regions `C' and `D' that predominantly appear to overlap with the composites and narrow-line AGNs but are gradually more distant from `A' and `B.' Region `E' captures a possible plume of points that also overlaps with composites and narrow-line AGNs but extend toward more quiescent galaxies (retired and passive). Finally, region `F' is the most distinct and primarily overlaps with the second grouping of retired and passive galaxies. From these selected regions, Figure \ref{fig:bl_all} show the stacked spectra sequentially from top to bottom. As a reference, the stacked spectrum of 2000 SDSS quasars from \citet{Berk+2001} (VB01) is shown in light blue. The latter represents the average SDSS quasar, and we will use it for visual comparison in Section~\ref{sec:analysis_interpret}.

\section{Analysis}\label{sec:analysis}
\subsection{Interpreting Stacked Spectra from UMAP Regions}\label{sec:analysis_interpret}

The 2D UMAP projection separates star-forming galaxies from retired and passive galaxies while placing composite and narrow-line galaxies as the bridge between the two. This suggests there are gradients of features as we see in Figure \ref{fig:topbot_all} of the sequences of stacked spectra. Starting from star-forming galaxies, stacked spectrum \#7 is comparatively brighter at wavelengths between 4000\angstrom\ and 4500\angstrom\ resulting in a blue color typical of young to intermediate age stellar populations. 

As we move from star-forming galaxies to composite and either retired or passive galaxies, the spectra become redder, indicating a progression from younger to older stars. Supporting this interpretation, we also see two notable spectral breaks: the Balmer break around 3641\angstrom, and the 4000\angstrom\ break \citep{Balogh+1999}. The Balmer break is most pronounced in intermediate-age stellar populations (<1~Gyr) dominated by late B and A stars and becomes less prominent as the age of the stars increases \citep{Bruzual+2003,GonzalezDelgado+2005}. In contrast, the 4000\angstrom\ break becomes stronger in spectra dominated by older stellar populations (>1~Gyr), and we can see that it becomes progressively stronger toward the retired/passive galaxies loci (boxes 0, 1, and 2 of each sequence in dark blue, purple, and black lines of Figure~\ref{fig:topbot_all}).

On average, we see stronger emission lines with star-forming galaxies (Figure \ref{fig:topbot_ha}), but composite and narrow-line AGN galaxies influence \hb\ and \ha\ especially strongly in spectrum \#3 (yellow solid line). The difference between the solid and dashed line in spectra \#3 are due to the selection of two different galaxy types; one sample predominantly consists of AGNs (box 3 from the top sequence) while the other sample resides in a retired and passive region projected by UMAP (box 3 from the bottom sequence). 

Next, we compare the sequences to each other by searching for differences between the spectra plotted with solid versus dashed lines (Figure \ref{fig:topbot_all}). The most substantial difference between the two sequences is the continuum color below 3500\angstrom\ in the sense that the spectra from the bottom sequence are systematically redder than those from the top sequence. This systematic difference may be due to a varying dust attenuation between galaxies in the top versus the bottom sequence. 
For boxes 0, 1, and 2, the bottom sequence (dashed line) consists of more passive galaxies than the top sequence, and the top sequence may consist of more retired galaxies. The separation of these two classes would normally imply stronger emission lines in the bottom sequence but we see the opposite due to the increased presence of active (AGN or LINER) galaxies interwoven with the top sequence. UMAP may be differentiating retired and passive galaxies on a different line from our definition based on EW(\ha), such as \nii, or the model may rely on the color of the continuum outside of emission line regions. We can indeed see some differences for boxes 0 and 1 around 5500-6200\AA.

Focusing on broad-line galaxies only, Figure \ref{fig:bl_all} shows the stacked spectrum `A' is nearly identical to the \citet{Berk+2001} stack except for the relative strength of the \oiii\ doublet. Other than the broad Balmer lines, spectrum `A' shows a strongly blue-colored spectrum and notable \feii\ lines around 4600\angstrom\ and 6800\angstrom. As the sequences moves gradually away from `A' and toward regions `B,' `C,' and `D,' we see increasingly redder slopes together with the appearance of host galaxies' absorption lines (\caii, H \& K lines at 3935\angstrom, 3970\angstrom). Previous studies \citep[e.g.,][]{Gordon+2003,Ross+2015,Fawcett+2022} propose that the difference in spectral slope from regular quasars and “Red quasars" can be explained by foreground dust (present in the galaxies hosting the quasars). Going further toward region `E,' we now see that for an equally red spectrum relative to `D,' the broad lines are less prominent with respect to continuum, suggesting either a smaller fraction of galaxies with a broad-line AGN or that those objects have a less luminous broad-line AGN in proportion to the host galaxy stellar light. Spectrum `F' is an outlier in terms of completely lacking broad Balmer lines and is instead dominated by old stars with strong \nii\ emission lines. This can be explained by the definition of the BROADLINE label in the SDSS pipeline not being strictly applied to permitted lines and having a low linewidth threshold of 200 km/s. Indeed, it seems that the \nii\ lines may be sufficiently broadened to meet the 200 km/s criterion due to the deep potential well in these massive LINER-like galaxies. We conclude that region `F' from Figure \ref{fig:umap_bl} is dominated by LINER galaxies and not genuine broad-line AGNs.

\subsection{Comparing to Previous Works}\label{sec:analysis_similar_diff}
The PAE is expected to show improvement from the VAE on encoding the entire continuum spectrum because the model learns with the added mask and noise. Through our AE corner plot (Figure \ref{fig:corner}), we can see the compressed data separates labels well, but clear weaknesses like manageability and interpretability continue if we use the 10D compressed data. With the help of additional dimensionality reduction algorithms, multidimensional spaces are less daunting to interpret. We find that the model accurately reconstructs most SDSS spectra at 200 epochs AE training, but improvements can be made where broad-lines are present around \hb\ and \ha. Refer to Appendix~\ref{app:pae} for more details on the PAE reconstruction performance.

A broad review of deep learning applications for discovery on galaxy surveys points out current unsolved issues on anomaly detection efficiency from oversimplification, anomaly detection uncertainty from artifacts, and the lack of interpretability to physical laws from deep learning parameters \citep{Huertas-Company+2022}. To address these issues, we optimize our PAE with a Fully Connected architecture to preserve the global structure (continua trends) while learning local structure (emission and absorption lines). Additionally, the integration of GIS introduces a probabilistic nature useful for anomaly detection for artifacts and actual rare spectra. Detecting anomalies from artifacts is needed for further data pruning and may suggest an underlying property in the data set itself. On the last note, interpreting physical laws from the AE or UMAP is still difficult, but with the use of stacked spectra instead of synthetic spectra, we see strong, averaged trends between classes that correspond to what we know as far as consistent.

On the note of general SDSS spectral patterns, \citet{Burnet+2021} found four general groups of classification for their SDSS DR7 sample: spectra with flat continua, negative slope continua, positive slope continua, and very blue continua with many emission lines. Comparing our GIS-selected spectra and UMAP-selected spectra to \citet{Burnet+2021}, we see the same general trends of spectral continua slope and an isolation of very blue continua. Whether or not the algorithms and neural networks are weighing these characteristics the most needs further investigation. Comparing our stacked spectra to \citet{Dobos+2012}, they produced high quality, co-added spectra of various classes including star-forming, passive, LINER, red, and blue galaxies, and we see similar trends in many respects like emission line strengths around \ha\ and prominent Balmer and 4000~\AA\ breaks. 

Altogether, UMAP is able to retain global structure of spectra while classifying galaxies in different spaces, and it is useful to implement a versatile dimensionality reduction algorithm that complements the reconstruction of the decoder. More information on applications and analysis is in Appendix~\ref{app:pae} on reconstruction accuracy.

\section{Conclusions and Future Work}\label{sec:conclusion}

Since the publication of BPT in 1981, the diagnostic diagram has been a popular tool for defining classifications of astronomical objects based on their primary source of ionization. The diagram is a great starting point to utilize spectroscopic data for extracting information, but it neglects most of the spectrum while introducing bias to machine learning models if only the four emission lines listed are given as input. Therefore, we propose an alternative or complement to analyze current and upcoming large spectroscopic surveys. 

Using a fixed rest frame wavelength grid of 1000 elements from 3388\angstrom\ to 8318\angstrom\ as our input, the AE is able to encode and decode the entire spectrum accurately with room for improvement at \ha\ and \hb\ emission lines due to their broad-line nature for Type 1 quasars and AGNs. To improve spectral reconstruction performance, analyzing the encoded compressed data to determining outliers and general data set trends for further data preprocessing can be done with the help of dimensionality reduction algorithms. Our integration of an AE with UMAP accomplished exactly these two goals and more.

In the future, we can increase the input data spectral resolution by doubling or tripling the number of elements, and the increased resolution may reveal more detailed information from the emission line profile such as linewidths or shapes. Optimizing the PAE may include a loss function analysis with Gaussian Kernel loss and Mixture Density Network loss as comparisons to the reduced \(\chi^{2}\). To help the AE training converge to a global minimum faster, we can change the type of loss function between the encoder and decoder and implement other neural network architectures. However, \citet{Boehm+2022} found no improvement of a Convolutional Neural Network architecture over a FC architecture. Being fully connected, the network can learn to optimize both local and global features.

For GIS, we manually set the training epoch stopping point at 200 epochs due to the nature of the algorithm to run until a global minimum is reached. This results in any marginal improvement to continue training, which is claimed to be stable at thousands of epochs. A recommended starting point for GIS is to set it to 500 epochs or at least the number of epochs of the AE. Still, most outliers determined by GIS's probability distribution are quasars, and the next step is to determine if GIS is separating broad-line galaxies based on lower statistical sample size or the difference in spectral quality and features. More sophisticated labels can be considered for a Conditional GIS model that trains on the labels as well.

For UMAP, we shortly explored 3D and interactive scatter plots to investigate mixed-class islands and the internal structure of the data set. With the 2D projection, stacked spectra show clear trends including emission line strength, blue or red continuum slopes at short wavelengths, and different feature contributions corresponding to the source ionization, stellar ages, and the presence of dust obscuration. Information lost in 2D projections include where islands branch from, the density distribution of the data set, and clear demarcation lines between transitional classes. 

Looking forward, complementary clustering algorithms and neural networks are needed for ease of use and increasing performance.
To connect UMAP to spectral reconstruction and data set preprocessing, additional UMAP options and work include hyperparameter fine-tuning, k-Nearest Neighbor Graph Neural Networks, and outlier detection to crossmatch with GIS.

Finally, outlier or misclassified spectra should be further examined either visually or with spectral fitting; some spectral fitting models include stellar synthesis population models, photoionization models, or spectral fitting models that decompose galaxy and AGN contributions. Analyzing our stacked spectra with spectral fitting may help reveal and quantify physical trends related to stellar population properties and/or the presence of an active black hole. On the whole, PAE with dimensionality reduction algorithms show promising results in solving spectra regression and classification problems, and information from rich data sets provided by spectroscopic surveys can be extracted at a greater potential. A prospective application of PAE and UMAP is for the Dark Energy Spectroscopic Instrument \citep[DESI;][]{DESI+2022} survey that is currently underway and will obtain more than 40 million spectra of galaxies and quasars \citep{DESI+2016a}.

\acknowledgements
This research uses services and data provided by the Astro Data Lab at NSF’s National Optical-Infrared Astronomy Research Laboratory. NOIRLab is operated by the Association of Universities for Research in Astronomy (AURA), Inc. under a cooperative agreement with the National Science Foundation. This research also uses services or data provided by the National Energy Research Scientific Computing Center. NERSC is operated by Lawrence Berkeley National Laboratory for the United States Department of Energy Office of Science. F. Pat thanks collaborators at University of California, Berkeley, Lawrence Berkeley National Laboratory, and NSF's NOIRLab for the guidance and discussions. Thank you Rebecca Lipson and Gurtina Besla at the University of Arizona's TIMESTEP program for connecting and supporting underrepresented groups to Tucson employers.

Funding for the SDSS and SDSS-II has been provided by the Alfred P. Sloan Foundation, the Participating Institutions, the National Science Foundation, the U.S. Department of Energy, the National Aeronautics and Space Administration, the Japanese Monbukagakusho, the Max Planck Society, and the Higher Education Funding Council for England. Funding for SDSS-III has been provided by the Alfred P. Sloan Foundation, the Participating Institutions, the National Science Foundation, and the U.S. Department of Energy Office of Science. Funding for the Sloan Digital Sky Survey IV has been provided by the Alfred P. Sloan Foundation, the U.S. Department of Energy Office of Science, and the Participating Institutions. SDSS-IV acknowledges support and resources from the Center for High Performance Computing at the University of Utah. The SDSS website is www.sdss.org.

SDSS-IV is managed by the Astrophysical Research Consortium for the Participating Institutions of the SDSS Collaboration including the Brazilian Participation Group, the Carnegie Institution for Science, Carnegie Mellon University, Center for Astrophysics | Harvard \& Smithsonian, the Chilean Participation Group, the French Participation Group, Instituto de Astrof\'isica de Canarias, The Johns Hopkins University, Kavli Institute for the Physics and Mathematics of the Universe (IPMU) / University of Tokyo, the Korean Participation Group, Lawrence Berkeley National Laboratory, Leibniz Institut f\"ur Astrophysik Potsdam (AIP),  Max-Planck-Institut f\"ur Astronomie (MPIA Heidelberg), 
Max-Planck-Institut f\"ur Astrophysik (MPA Garching), Max-Planck-Institut f\"ur Extraterrestrische Physik (MPE), National Astronomical Observatories of China, New Mexico State University, New York University, University of Notre Dame, Observat\'ario Nacional / MCTI, The Ohio State University, Pennsylvania State University, Shanghai Astronomical Observatory, United Kingdom Participation Group, Universidad Nacional Aut\'onoma de M\'exico, University of Arizona, University of Colorado Boulder, University of Oxford, University of Portsmouth, University of Utah, University of Virginia, University of Washington, University of Wisconsin, Vanderbilt University, and Yale University.

\appendix
\section{PAE Performance Evaluation}\label{app:pae}

We evaluate the performance by considering the data compression, the reduction in the loss function as the model is trained, and the reconstruction error. The Fully Connected architecture depicted in Figure \ref{fig:nn_info} shows the drastic data compression process in three layers, from the initial 1000 to the final 10 parameters for each spectrum.

\articlefigure[scale=0.7]{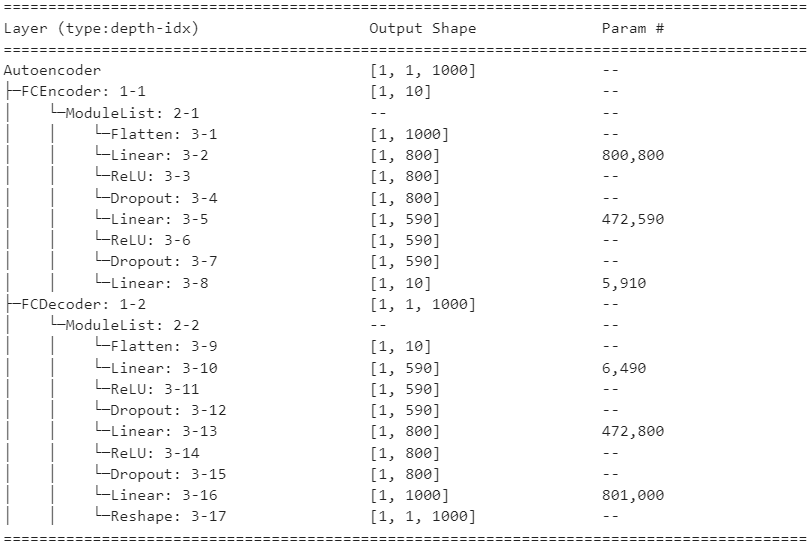}{fig:nn_info}
{Architecture of the PAE showing the tensor dimensions and purpose of each layer.}

From Figure \ref{fig:loss}, the AE training has not reached a global minimum yet and has not overtrained on the data set. Future work could include either training for more epochs to converge to a loss value closer to 1.0 or introducing a combination of loss functions during training. However, if theorized improvements on the neural network architecture and hyperparameter fine-tuning does not improve training efficiency and performance, improving data set preprocessing would be the next step. For example, B\"ohm et al. (in preparation) also found that an AE with the same loss function trained on a similar SDSS DR16 sample does not reach 1.0 and argue that this indicates that the measurement uncertainties are underestimated. They found that increasing the variance by 5-10\% allows the loss function to reach closer to 1.0 but does not solve the strong variation of the $\chi^2$ along the spectral direction so they opt to keep the original flux variance values. A detailed investigation of the SDSS spectra uncertainties is beyond the scope of this exploratory work but could be relevant for future studies.

\articlefigure[scale=0.5]{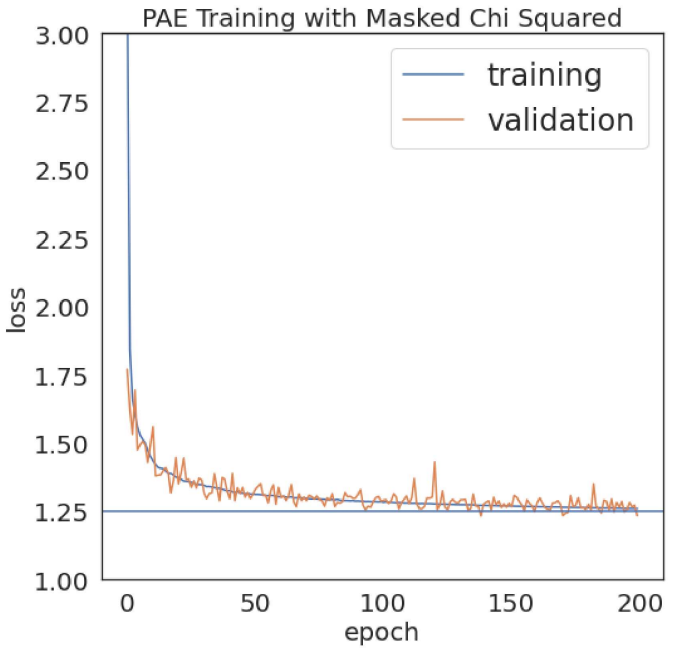}{fig:loss}
{\(\chi^{2}\) loss as a function of epoch (training iteration) to show the convergence and training progress of the AE. A horizontal line at 1.25 is plotted for comparison purposes.}

With the 200 epoch AE, the encoder and decoder are self-sufficient with a \(\chi^{2}\) loss value of \(1.26\). The 10D compressed data (Figure \ref{fig:corner}) shows some separation between star-forming, narrow-line AGN, and retired and passive galaxies. We chose to use 10D because \citet{Portillo+2020} found marginal improvement in spectral reconstruction accuracy from six parameters to ten. While the separations between classes are fuzzy, and it is difficult to interpret the compressed data given the high dimensionality, the scatter plots show that the AE is promising as we see hints of separation between the labels, which were not used in the training. This could mean that the model is picking up the same spectral regions used by the BPT and WHAN diagrams or that the labels are a good starting point to classify the spectra of galaxies. However, we caution that the corner plots of the AE embedded space should not be overinterpreted. There are many ways to compress data down and populate a 10D space, and the results are dependent on the initial random seed. 

\articlefigure[scale=0.28]{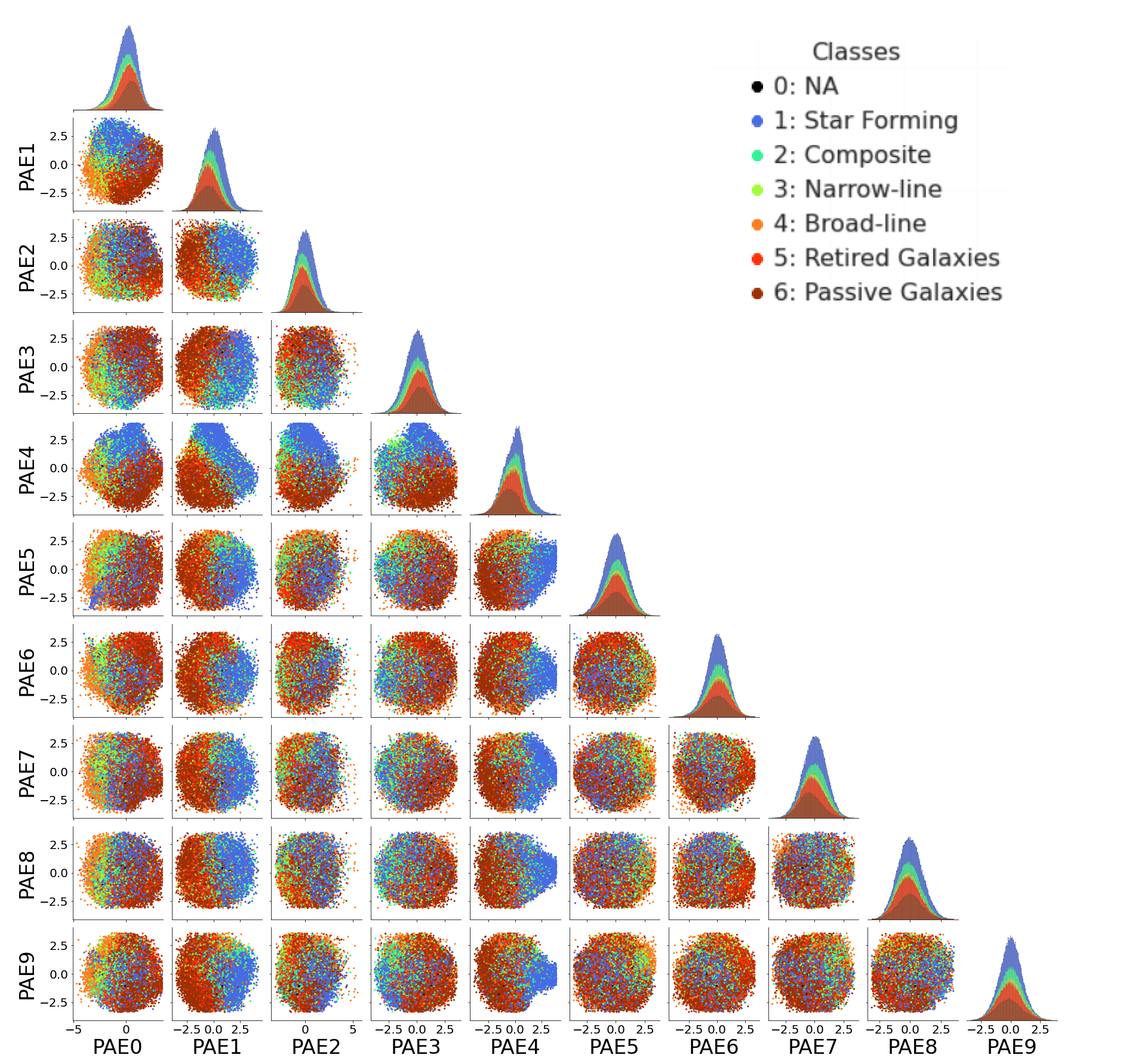}{fig:corner}
{Corner plot of the 10D compressed data with all plots trimmed from the top and bottom \(0.01\%\) outliers. Parameter 0 (PAE0) scatter plots are further zoomed in to better show the bulk of the population, which excludes a few broad-line (orange points) galaxies that are likely quasars. There are forty-five 2D scatter plots of arbitrary units displaying the spatial occupation of each class on 2D slices and ten 1D histograms showing the spread differentiated by label. Each dot is color-coded by the classification label as listed in the legend.}

The normalizing flow step is therefore useful as it mitigates against this dependency of the encoded latent space on the random seed and the result of the density estimator is more reproducible \citep{Boehm+2022}. In details, by using the 10D compressed data as input, SINF-GIS treats each parameter as a 1D slice such that an Optimal Transport (OT) solution is assumed to exist between each parameter. GIS estimates the probability density of a complex distribution by Gaussianizing it (bijective transformation) assuming an OT solution and minimizing the Maximum K-sliced p-Wasserstein Distance \citep[max K-SWD;][]{Dai+2020}, which is a distance function between probability distributions. After training for 200 epochs, Figure \ref{fig:gis_dens} shows the GIS probability distribution of the reduced compressed data. The shape of the distribution deviates from a Gaussian. Notably, it presents a tail toward low probability values that can be used to search for outliers or rare spectra. Conversely, selecting from the peak or high probability cases will reveal the most common types of galaxy spectra.

\articlefigure[scale=0.5]{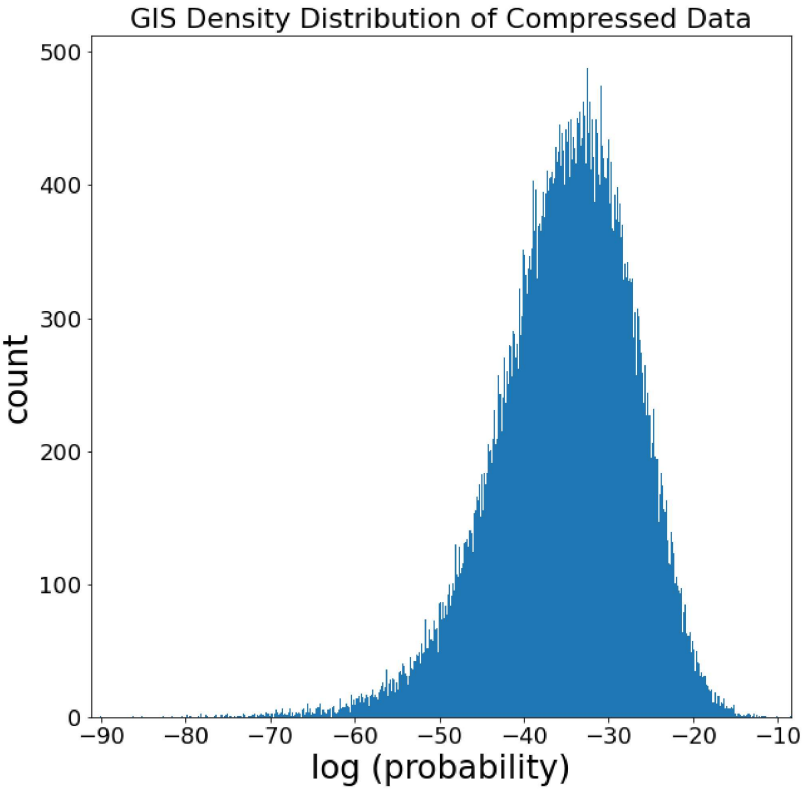}{fig:gis_dens}
{After training 200 epochs, we evaluate the density distribution of the compressed data with GIS to produce a histogram. The distribution is shown for the training set.}

From the density distribution produced by the normalizing flow, we can select data points to examine their spectra. Taking the 16 lowest \(\mathrm{log(probability)}\) samples show that the rarest spectra are mostly broad-line galaxies and a few retired galaxies (Figure \ref{fig:rare_spec}). An interesting, rare spectrum is in the third row and 1st column; the spectrum strongly resembles a BL Lac, which is a subtype of blazars with a featureless continuum dominated by non-thermal emission as indicated by \citet{deMenezes+2019} in their work on selecting and classifying the optical spectra of blazar candidates \citep[also see review by ][]{Falomo+2014}. This object was labeled as a passive galaxy due to the absence of detectable emission lines (Figure~\ref{fig:rare_spec}). The current classification scheme does not include a blazar category, but BL Lacs can be differentiated from inactive (i.e., non-AGN) passive galaxies by using their optical color information.

From the same GIS probability distribution, we randomly sample 16 reconstructed spectra to show the most common types of objects. Their reconstructed spectra (Figure \ref{fig:common_spec}) show two dominant types: spectra with prominent narrow-lines typical of star-forming galaxies, and spectra with mostly absorption lines that are typical of inactive passive galaxies. These are indeed similar to representative spectra from an atlas of templates based on low-redshift SDSS galaxies \citep{Dobos+2012}.

\clearpage
\articlefigure[scale=0.5]{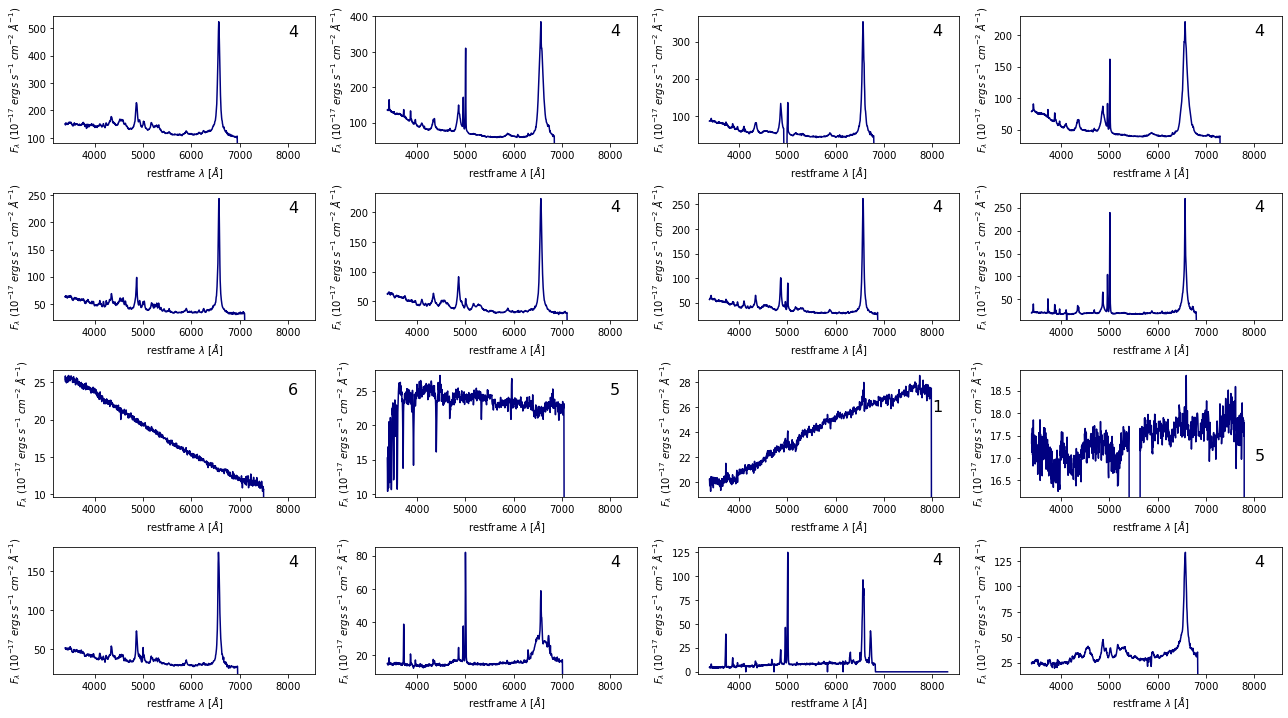}{fig:rare_spec}
{Taking the 16 lowest \(\mathrm{log(probability)}\) samples from Figure \ref{fig:gis_dens}, the rarest spectra are plotted with their labels on the upper right. Refer to Figure \ref{fig:umap_topbot} for the legend of class names. Masked regions appear as zero flux.}

\articlefigure[scale=0.5]{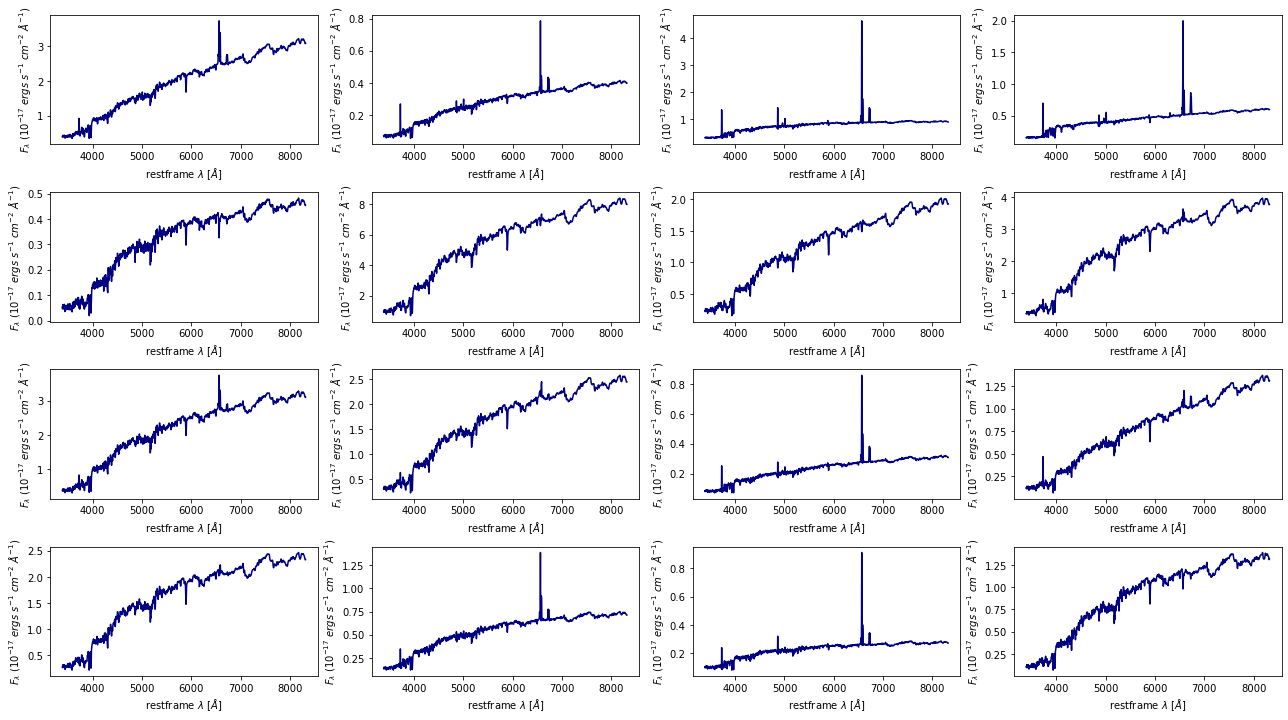}{fig:common_spec}
{Taking a sample of 16 spectra from Figure \ref{fig:gis_dens}, common reconstructed spectra show mostly two dominant populations: red spectra with absorption lines typical of passive galaxies, and star-forming galaxy spectra with both absorption lines and narrow emission lines.}

The reconstructed spectra, obtained via the decoder portion of the AE, have the advantages of being noiseless and of interpolating over masked regions from the original spectrum. Using the compressed data as the input to the decoder, the reconstruction performance is measured by:
\begin{equation}\label{eq:chi_2}
{\chi}^2=\mathrm{\langle \frac{1}{F^{mask}} \rangle \cdot \frac{1}{1000}\sum_{pixel=0}^{999} (F_{pixel}^{recon}-F_{pixel}^{data})^2\cdot{F_{pixel}^{noise}}\cdot{F_{pixel}^{mask}}}
\end{equation}
where $\mathrm{F_{pixel}^{recon}}$ is the reconstructed spectrum flux, $\mathrm{F_{pixel}^{data}}$ is the measured flux, $\mathrm{F_{pixel}^{noise}}$ is the measurement error, and $\mathrm{F_{pixel}^{mask}}$ is the binary mask.

\begin{figure}[!htb]
    
    \begin{subfigure}[t]{\textwidth}
    \centering
        \includegraphics[width=\linewidth]{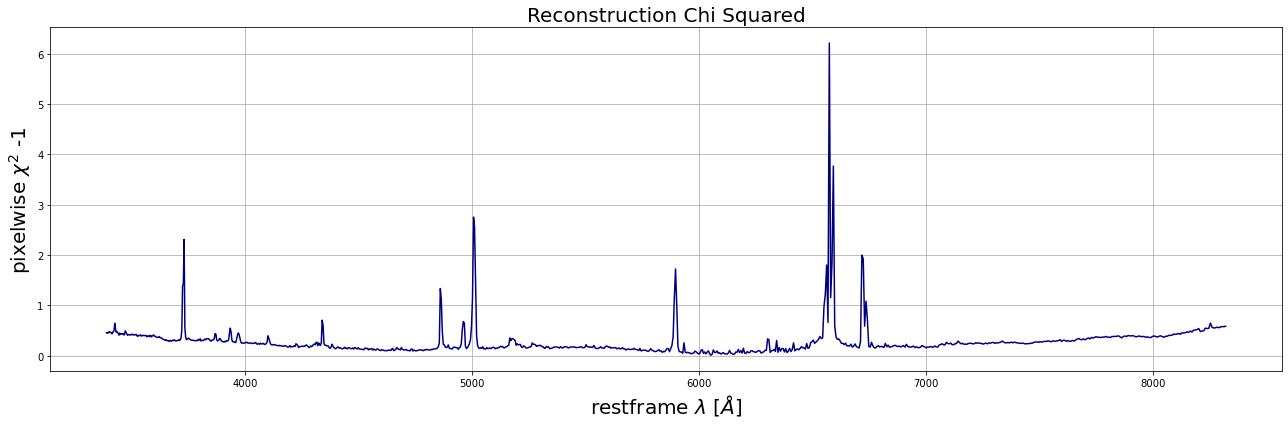} 
        \caption{Reduced pixelwise \(\chi^{2}-1\) of validation set reconstruction as a function of wavelength (\angstrom)} \label{fig:chi_squared}
    \end{subfigure}
    \vspace{0.25cm}
    
    \centering
    \begin{subfigure}[t]{0.48\textwidth}
        \centering
        \includegraphics[width=\linewidth]{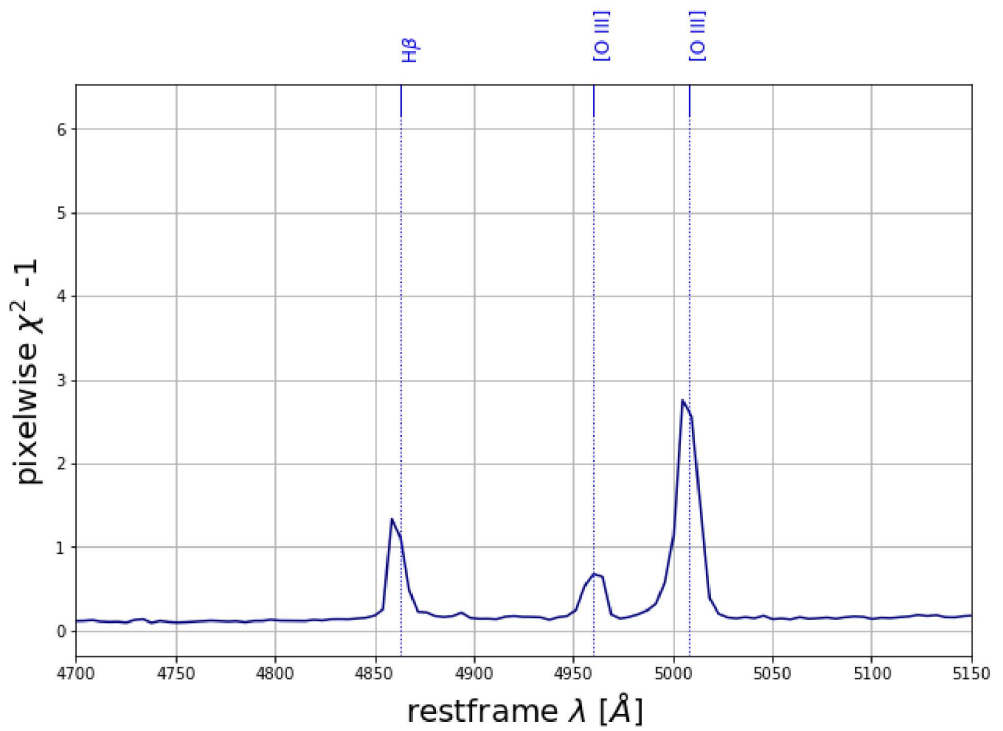} 
        \caption{Zoomed in reduced pixelwise \(\chi^{2}-1\) around \hb\ and \oiii, as labeled.} \label{fig:chi_squared_hb}
    \end{subfigure}
    \hfill
    \begin{subfigure}[t]{0.48\textwidth}
        \centering
        \includegraphics[width=\linewidth]{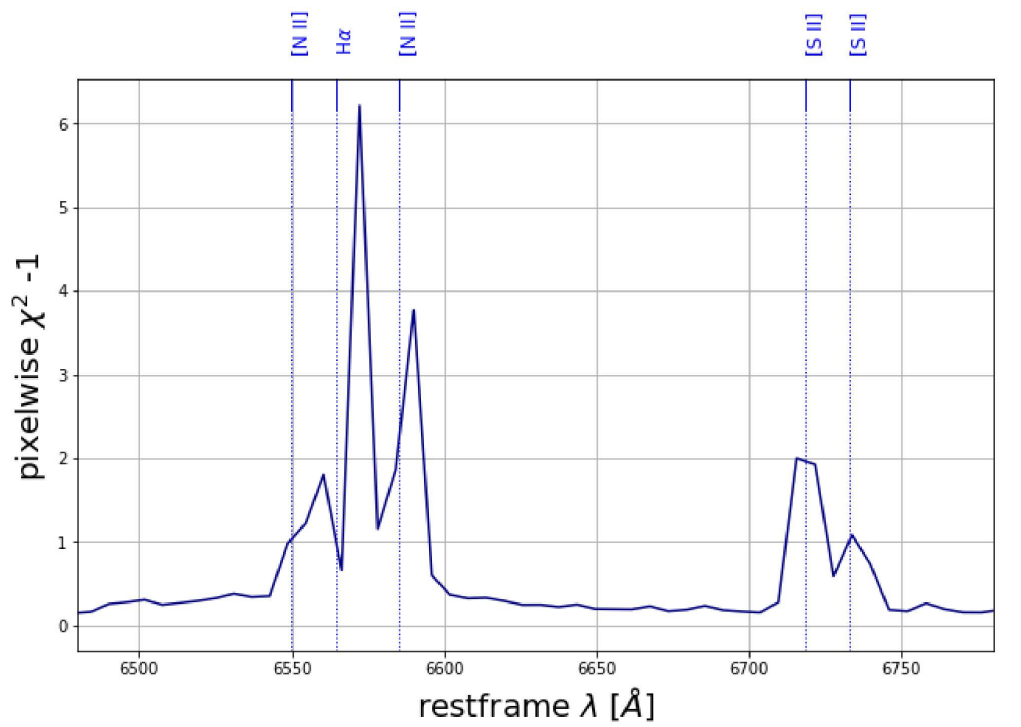} 
        \caption{Zoomed in reduced pixelwise \(\chi^{2}-1\) around \ha, \nii\ and \sii, as labeled.} \label{fig:chi_squared_ha}
    \end{subfigure}
    
    \caption{Using the validation set, the reduced pixelwise \(\chi^{2}-1\) as a function of wavelength shows lower reconstruction accuracy around \ha\ and \nii. Zooming in around \ha\ (Figure~\ref{fig:chi_squared_ha}) shows a skew around \ha\ not seen at the location of other emission lines.}
\end{figure}

Figure \ref{fig:chi_squared} shows the pixelwise $\chi^2$ for the validation set. We subtract a value of unity so a perfect reconstruction would have a value of zero. Focusing on the continuum, the reconstruction achieves $\chi^2-1<0.5$ over most of the spectral range with elevated values toward both ends of the range. In general, more galaxies have zeroed out mask near the edges of the spectrum if they do not cover the full wavelength grid when de-redshifted to the rest-frame. Accordingly, the data are not constraining the model as strongly there (and if it is also the end of the observed spectrum, it risks being noisier), which we see with the rising \(\chi^{2}\) toward both the blue end and the red end. Regions with the highest \(\chi^{2}\) coincide with the expected locations of strong emission lines. Namely, it reaches the highest values around the \ha\ and \nii\ complex. These spectral features can vary widely between quasars and inactive galaxies, and even among quasars there can be substantial variations in emission line strengths and profiles. 
Lastly, we examine the \(\chi^{2}\) around the regions spanning strong emission lines (Figures~\ref{fig:chi_squared_hb} and \ref{fig:chi_squared_ha}). Zooming around \ha\ (Figure \ref{fig:chi_squared_ha}), the emission lines seem to be off-centered showing broad-line spectral features affecting the model's learning and reconstruction performance. The reason for this apparent offset is not fully clear. It could be due to uncertain redshifts (especially in broad line cases) or simply because of the rare nature of the most extreme emission line strengths and profiles hindering a complete training of the model. Future work could include attempting to train separately the quasars from the remainder of the sample and/or using higher-resolution input spectra to include emission line profile information.

As an interesting comparison, \citet{Ferreras+2022} presented an approach to quantify the amount of information encoded in galaxy spectra via a measurement of the entropy as a function of wavelength. For optical SDSS galaxy spectra, they found a high entropy around \ha\ and \hb\ for star-forming and AGN galaxies, where the \(\chi^{2}\) show peaks. Additionally, they found that the most entropy is contained around the Balmer and 4000~\AA\ breaks, which can correlate with stellar age; however, we obtain a relatively low \(\chi^{2}\) around the Balmer and 4000~\AA\ breaks. Thus, spectral areas with high entropy are not a one-to-one mapping with elevated \(\chi^{2}\). We deduce from the relatively low \(\chi^{2}\) that the model is well trained to pick up on those information-rich spectral breaks. 

We trained all models presented here on the National Energy Research Scientific Computing Center (NERSC) Perlmutter GPU; as a general standard, subsequent attempts to exactly replicate the work discussed in this paper should be based on the same hardware and software specifications\footnote{https://docs.nersc.gov/systems/perlmutter/}. We found slight differences when using CPU versus GPU, which are likely due to differences in the compilers. They can result in slightly different numerical values on the UMAP projections without however affecting the overall trends and topology. We used Jupyter Notebooks \citep{Kluyver+2016} at both NERSC and the NOIRLab Astro Data Lab \citep{Juneau+2021} for the bulk of the analysis. 


\bibliography{refs}

\begin{thebibliography}{}
\expandafter\ifx\csname natexlab\endcsname\relax\def\natexlab#1{#1}\fi
\providecommand{\url}[1]{\href{#1}{#1}}
\providecommand{\dodoi}[1]{doi:~\href{http://doi.org/#1}{\nolinkurl{#1}}}
\providecommand{\doeprint}[1]{\href{http://ascl.net/#1}{\nolinkurl{http://ascl.net/#1}}}
\providecommand{\doarXiv}[1]{\href{https://arxiv.org/abs/#1}{\nolinkurl{https://arxiv.org/abs/#1}}}

\bibitem[{{Ahumada} {et~al.}(2020){Ahumada}, {Prieto}, {Almeida}, {Anders},
  {Anderson}, \& {et al.}}]{Ahumada+2020}
{Ahumada}, R., {Prieto}, C.~A., {Almeida}, A., {et~al.} 2020, \apjs, 249, 3,
  \dodoi{10.3847/1538-4365/ab929e}

\bibitem[{{Baldwin} {et~al.}(1981){Baldwin}, {Phillips}, \&
  {Terlevich}}]{Baldwin+1981}
{Baldwin}, J.~A., {Phillips}, M.~M., \& {Terlevich}, R. 1981, \pasp, 93, 5,
  \dodoi{10.1086/130766}

\bibitem[{{Balogh} {et~al.}(1999){Balogh}, {Morris}, {Yee}, {Carlberg}, \&
  {Ellingson}}]{Balogh+1999}
{Balogh}, M.~L., {Morris}, S.~L., {Yee}, H.~K.~C., {Carlberg}, R.~G., \&
  {Ellingson}, E. 1999, \apj, 527, 54, \dodoi{10.1086/308056}

\bibitem[{B\"ohm \& Seljak(2022)}]{Boehm+2022}
B\"ohm, V.~M., \& Seljak, U. 2022, Transactions on Machine Learning Research.
\newblock \url{https://openreview.net/forum?id=AEoYjvjKVA}

\bibitem[{{Bolton} {et~al.}(2012){Bolton}, {Schlegel}, {Aubourg}, {Bailey},
  {Bhardwaj}, {Brownstein}, {Burles}, {Chen}, {Dawson}, {Eisenstein}, {Gunn},
  {Knapp}, {Loomis}, {Lupton}, {Maraston}, {Muna}, {Myers}, {Olmstead},
  {Padmanabhan}, {P{\^a}ris}, {Percival}, {Petitjean}, {Rockosi}, {Ross},
  {Schneider}, {Shu}, {Strauss}, {Thomas}, {Tremonti}, {Wake}, {Weaver}, \&
  {Wood-Vasey}}]{Bolton+2012}
{Bolton}, A.~S., {Schlegel}, D.~J., {Aubourg}, {\'E}., {et~al.} 2012, \aj, 144,
  144, \dodoi{10.1088/0004-6256/144/5/144}

\bibitem[{{Bruzual} \& {Charlot}(2003)}]{Bruzual+2003}
{Bruzual}, G., \& {Charlot}, S. 2003, \mnras, 344, 1000,
  \dodoi{10.1046/j.1365-8711.2003.06897.x}

\bibitem[{{Cid Fernandes} {et~al.}(2011){Cid Fernandes}, {Stasi{\'n}ska},
  {Mateus}, \& {Vale Asari}}]{CidFernandes+2011}
{Cid Fernandes}, R., {Stasi{\'n}ska}, G., {Mateus}, A., \& {Vale Asari}, N.
  2011, \mnras, 413, 1687, \dodoi{10.1111/j.1365-2966.2011.18244.x}

\bibitem[{{Clarke} {et~al.}(2020){Clarke}, {Scaife}, {Greenhalgh}, \&
  {Griguta}}]{Clarke+2020}
{Clarke}, A.~O., {Scaife}, A.~M.~M., {Greenhalgh}, R., \& {Griguta}, V. 2020,
  \aap, 639, A84, \dodoi{10.1051/0004-6361/201936770}

\bibitem[{{Dai} \& {Seljak}(2020)}]{Dai+2020}
{Dai}, B., \& {Seljak}, U. 2020, arXiv e-prints, arXiv:2007.00674.
\newblock \doarXiv{2007.00674}

\bibitem[{{de Menezes} {et~al.}(2019){de Menezes}, {Pe{\~n}a-Herazo},
  {Marchesini}, {D'Abrusco}, {Masetti}, {Nemmen}, {Massaro}, {Ricci},
  {Landoni}, {Paggi}, \& {Smith}}]{deMenezes+2019}
{de Menezes}, R., {Pe{\~n}a-Herazo}, H.~A., {Marchesini}, E.~J., {et~al.} 2019,
  \aap, 630, A55, \dodoi{10.1051/0004-6361/201936195}

\bibitem[{{DESI Collaboration} {et~al.}(2016){DESI Collaboration}, {Aghamousa},
  {Aguilar}, {Ahlen}, {Alam}, {Allen}, {Allende Prieto}, {Annis}, {Bailey},
  {Balland}, {Ballester}, {Baltay}, {Beaufore}, {Bebek}, {Beers}, {Bell},
  {Bernal}, {Besuner}, {Beutler}, {Blake}, {Bleuler}, {Blomqvist}, {Blum},
  {Bolton}, {Briceno}, {Brooks}, {Brownstein}, {Buckley-Geer}, {Burden},
  {Burtin}, {Busca}, {Cahn}, {Cai}, {Cardiel-Sas}, {Carlberg}, {Carton},
  {Casas}, {Castander}, {Cervantes-Cota}, {Claybaugh}, {Close}, {Coker},
  {Cole}, {Comparat}, {Cooper}, {Cousinou}, {Crocce}, {Cuby}, {Cunningham},
  {Davis}, {Dawson}, {de la Macorra}, {De Vicente}, {Delubac}, {Derwent},
  {Dey}, {Dhungana}, {Ding}, {Doel}, {Duan}, {Ealet}, {Edelstein},
  {Eftekharzadeh}, {Eisenstein}, {Elliott}, {Escoffier}, {Evatt}, {Fagrelius},
  {Fan}, {Fanning}, {Farahi}, {Farihi}, {Favole}, {Feng}, {Fernandez},
  {Findlay}, {Finkbeiner}, {Fitzpatrick}, {Flaugher}, {Flender}, {Font-Ribera},
  {Forero-Romero}, {Fosalba}, {Frenk}, {Fumagalli}, {Gaensicke}, {Gallo},
  {Garcia-Bellido}, {Gaztanaga}, {Pietro Gentile Fusillo}, {Gerard},
  {Gershkovich}, {Giannantonio}, {Gillet}, {Gonzalez-de-Rivera},
  {Gonzalez-Perez}, {Gott}, {Graur}, {Gutierrez}, {Guy}, {Habib}, {Heetderks},
  {Heetderks}, {Heitmann}, {Hellwing}, {Herrera}, {Ho}, {Holland}, {Honscheid},
  {Huff}, {Hutchinson}, {Huterer}, {Hwang}, {Illa Laguna}, {Ishikawa},
  {Jacobs}, {Jeffrey}, {Jelinsky}, {Jennings}, {Jiang}, {Jimenez}, {Johnson},
  {Joyce}, {Jullo}, {Juneau}, {Kama}, {Karcher}, {Karkar}, {Kehoe}, {Kennamer},
  {Kent}, {Kilbinger}, {Kim}, {Kirkby}, {Kisner}, {Kitanidis}, {Kneib},
  {Koposov}, {Kovacs}, {Koyama}, {Kremin}, {Kron}, {Kronig}, {Kueter-Young},
  {Lacey}, {Lafever}, {Lahav}, {Lambert}, {Lampton}, {Landriau}, {Lang},
  {Lauer}, {Le Goff}, {Le Guillou}, {Le Van Suu}, {Lee}, {Lee}, {Leitner},
  {Lesser}, {Levi}, {L'Huillier}, {Li}, {Liang}, {Lin}, {Linder}, {Loebman},
  {Luki{\'c}}, {Ma}, {MacCrann}, {Magneville}, {Makarem}, {Manera}, {Manser},
  {Marshall}, {Martini}, {Massey}, {Matheson}, {McCauley}, {McDonald},
  {McGreer}, {Meisner}, {Metcalfe}, {Miller}, {Miquel}, {Moustakas}, {Myers},
  {Naik}, {Newman}, {Nichol}, {Nicola}, {Nicolati da Costa}, {Nie}, {Niz},
  {Norberg}, {Nord}, {Norman}, {Nugent}, {O'Brien}, {Oh}, {Olsen}, {Padilla},
  {Padmanabhan}, {Padmanabhan}, {Palanque-Delabrouille}, {Palmese},
  {Pappalardo}, {P{\^a}ris}, {Park}, {Patej}, {Peacock}, {Peiris}, {Peng},
  {Percival}, {Perruchot}, {Pieri}, {Pogge}, {Pollack}, {Poppett}, {Prada},
  {Prakash}, {Probst}, {Rabinowitz}, {Raichoor}, {Ree}, {Refregier}, {Regal},
  {Reid}, {Reil}, {Rezaie}, {Rockosi}, {Roe}, {Ronayette}, {Roodman}, {Ross},
  {Ross}, {Rossi}, {Rozo}, {Ruhlmann-Kleider}, {Rykoff}, {Sabiu}, {Samushia},
  {Sanchez}, {Sanchez}, {Schlegel}, {Schneider}, {Schubnell}, {Secroun},
  {Seljak}, {Seo}, {Serrano}, {Shafieloo}, {Shan}, {Sharples}, {Sholl},
  {Shourt}, {Silber}, {Silva}, {Sirk}, {Slosar}, {Smith}, {Smoot}, {Som},
  {Song}, {Sprayberry}, {Staten}, {Stefanik}, {Tarle}, {Sien Tie}, {Tinker},
  {Tojeiro}, {Valdes}, {Valenzuela}, {Valluri}, {Vargas-Magana}, {Verde},
  {Walker}, {Wang}, {Wang}, {Weaver}, {Weaverdyck}, {Wechsler}, {Weinberg},
  {White}, {Yang}, {Yeche}, {Zhang}, {Zhao}, {Zheng}, {Zhou}, {Zhou}, {Zhu},
  {Zou}, \& {Zu}}]{DESI+2016a}
{DESI Collaboration}, {Aghamousa}, A., {Aguilar}, J., {et~al.} 2016, arXiv
  e-prints, arXiv:1611.00036.
\newblock \doarXiv{1611.00036}

\bibitem[{{DESI Collaboration} {et~al.}(2022){DESI Collaboration}, {Abareshi},
  {Aguilar}, {Ahlen}, {Alam}, {Alexander}, {Alfarsy}, {Allen}, {Allende
  Prieto}, {Alves}, {Ameel}, {Armengaud}, {Asorey}, {Aviles}, {Bailey},
  {Balaguera-Antol{\'\i}nez}, {Ballester}, {Baltay}, {Bault}, {Beltran},
  {Benavides}, {BenZvi}, {Berti}, {Besuner}, {Beutler}, {Bianchi}, {Blake},
  {Blanc}, {Blum}, {Bolton}, {Bose}, {Bramall}, {Brieden}, {Brodzeller},
  {Brooks}, {Brownewell}, {Buckley-Geer}, {Cahn}, {Cai}, {Canning}, {Carnero
  Rosell}, {Carton}, {Casas}, {Castander}, {Cervantes-Cota}, {Chabanier},
  {Chaussidon}, {Chuang}, {Circosta}, {Cole}, {Cooper}, {da Costa}, {Cousinou},
  {Cuceu}, {Davis}, {Dawson}, {de la Cruz-Noriega}, {de la Macorra}, {de
  Mattia}, {Della Costa}, {Demmer}, {Derwent}, {Dey}, {Dey}, {Dhungana},
  {Ding}, {Dobson}, {Doel}, {Donald-McCann}, {Donaldson}, {Douglass}, {Duan},
  {Dunlop}, {Edelstein}, {Eftekharzadeh}, {Eisenstein}, {Enriquez-Vargas},
  {Escoffier}, {Evatt}, {Fagrelius}, {Fan}, {Fanning}, {Fawcett}, {Ferraro},
  {Ereza}, {Flaugher}, {Font-Ribera}, {Forero-Romero}, {Frenk}, {Fromenteau},
  {G{\"a}nsicke}, {Garcia-Quintero}, {Garrison}, {Gazta{\~n}aga}, {Gerardi},
  {Gil-Mar{\'\i}n}, {Gontcho}, {Gonzalez-Morales}, {Gonzalez-de-Rivera},
  {Gonzalez-Perez}, {Gordon}, {Graur}, {Green}, {Grove}, {Gruen}, {Gutierrez},
  {Guy}, {Hahn}, {Harris}, {Herrera}, {Herrera-Alcantar}, {Honscheid},
  {Howlett}, {Huterer}, {Ir{\v{s}}i{\v{c}}}, {Ishak}, {Jelinsky}, {Jiang},
  {Jimenez}, {Jing}, {Joyce}, {Jullo}, {Juneau}, {Kara{\c{c}}ayl{\i}},
  {Karamanis}, {Karcher}, {Karim}, {Kehoe}, {Kent}, {Kirkby}, {Kisner},
  {Kitaura}, {Koposov}, {Kov{\'a}cs}, {Kremin}, {Krolewski}, {L'Huillier},
  {Lahav}, {Lambert}, {Lamman}, {Lan}, {Landriau}, {Lane}, {Lang}, {Lange},
  {Lasker}, {Le Guillou}, {Leauthaud}, {Le Van Suu}, {Levi}, {Li},
  {Magneville}, {Manera}, {Manser}, {Marshall}, {McCollam}, {McDonald},
  {Meisner}, {Mezcua}, {Miller}, {Miquel}, {Montero-Camacho}, {Moon},
  {Martini}, {Meneses-Rizo}, {Moustakas}, {Mueller}, {Mu{\~n}oz-Guti{\'e}rrez},
  {Myers}, {Nadathur}, {Najita}, {Napolitano}, {Neilsen}, {Newman}, {Nie},
  {Ning}, {Niz}, {Norberg}, {Noriega}, {O'Brien}, {Obuljen},
  {Palanque-Delabrouille}, {Palmese}, {Zhiwei}, {Pappalardo}, {Peng},
  {Percival}, {Perruchot}, {Pogge}, {Poppett}, {Porredon}, {Prada},
  {Prochaska}, {Pucha}, {P{\'e}rez-Fern{\'a}ndez}, {P{\'e}rez-R{\'a}fols},
  {Rabinowitz}, {Raichoor}, {Ramirez-Solano}, {Ram{\'\i}rez-P{\'e}rez},
  {Ravoux}, {Reil}, {Rezaie}, {Rocher}, {Rockosi}, {Roe}, {Roodman}, {Ross},
  {Rossi}, {Ruggeri}, {Ruhlmann-Kleider}, {Sabiu}, {Safonova}, {Said},
  {Saintonge}, {Salas Catonga}, {Samushia}, {Sanchez}, {Saulder}, {Schaan},
  {Schlafly}, {Schlegel}, {Schmoll}, {Scholte}, {Schubnell}, {Secroun}, {Seo},
  {Serrano}, {Sharples}, {Sholl}, {Silber}, {Silva}, {Sirk}, {Siudek}, {Smith},
  {Sprayberry}, {Staten}, {Stupak}, {Tan}, {Tarl{\'e}}, {Sien Tie}, {Tojeiro},
  {Ure{\~n}a-L{\'o}pez}, {Valdes}, {Valenzuela}, {Valluri},
  {Vargas-Maga{\~n}a}, {Verde}, {Walther}, {Wang}, {Wang}, {Weaver},
  {Weaverdyck}, {Wechsler}, {Wilson}, {Yang}, {Yu}, {Yuan}, {Y{\`e}che},
  {Zhang}, {Zhang}, {Zhao}, {Zhou}, {Zhou}, {Zou}, {Zou}, {Zou}, \&
  {Zu}}]{DESI+2022}
{DESI Collaboration}, {Abareshi}, B., {Aguilar}, J., {et~al.} 2022, arXiv
  e-prints, arXiv:2205.10939.
\newblock \doarXiv{2205.10939}

\bibitem[{{Dobos} {et~al.}(2012){Dobos}, {Csabai}, {Yip}, {Budav{\'a}ri},
  {Wild}, \& {Szalay}}]{Dobos+2012}
{Dobos}, L., {Csabai}, I., {Yip}, C.-W., {et~al.} 2012, \mnras, 420, 1217,
  \dodoi{10.1111/j.1365-2966.2011.20109.x}

\bibitem[{{Falomo} {et~al.}(2014){Falomo}, {Pian}, \& {Treves}}]{Falomo+2014}
{Falomo}, R., {Pian}, E., \& {Treves}, A. 2014, \aapr, 22, 73,
  \dodoi{10.1007/s00159-014-0073-z}

\bibitem[{{Fawcett} {et~al.}(2022){Fawcett}, {Alexander}, {Rosario}, {Klindt},
  {Lusso}, {Morabito}, \& {Calistro Rivera}}]{Fawcett+2022}
{Fawcett}, V.~A., {Alexander}, D.~M., {Rosario}, D.~J., {et~al.} 2022, \mnras,
  513, 1254, \dodoi{10.1093/mnras/stac945}

\bibitem[{{Ferreras} {et~al.}(2022){Ferreras}, {Lahav}, {Somerville}, \&
  {Silk}}]{Ferreras+2022}
{Ferreras}, I., {Lahav}, O., {Somerville}, R.~S., \& {Silk}, J. 2022, arXiv
  e-prints, arXiv:2208.05489.
\newblock \doarXiv{2208.05489}

\bibitem[{{Fitzpatrick} {et~al.}(2014){Fitzpatrick}, {Olsen}, {Economou},
  {Stobie}, {Beers}, {Dickinson}, {Norris}, {Saha}, {Seaman}, {Silva},
  {Swaters}, {Thomas}, \& {Valdes}}]{Fitzpatrick+2014}
{Fitzpatrick}, M.~J., {Olsen}, K., {Economou}, F., {et~al.} 2014, in Society of
  Photo-Optical Instrumentation Engineers (SPIE) Conference Series, Vol. 9149,
  Observatory Operations: Strategies, Processes, and Systems V, ed. A.~B.
  {Peck}, C.~R. {Benn}, \& R.~L. {Seaman}, 91491T, \dodoi{10.1117/12.2057445}

\bibitem[{{Fraix-Burnet} {et~al.}(2021){Fraix-Burnet}, {Bouveyron}, \&
  {Moultaka}}]{Burnet+2021}
{Fraix-Burnet}, D., {Bouveyron}, C., \& {Moultaka}, J. 2021, \aap, 649, A53,
  \dodoi{10.1051/0004-6361/202040046}

\bibitem[{{Gonz{\'a}lez Delgado} {et~al.}(2005){Gonz{\'a}lez Delgado},
  {Cervi{\~n}o}, {Martins}, {Leitherer}, \&
  {Hauschildt}}]{GonzalezDelgado+2005}
{Gonz{\'a}lez Delgado}, R.~M., {Cervi{\~n}o}, M., {Martins}, L.~P.,
  {Leitherer}, C., \& {Hauschildt}, P.~H. 2005, \mnras, 357, 945,
  \dodoi{10.1111/j.1365-2966.2005.08692.x}

\bibitem[{{Huertas-Company} \& {Lanusse}(2022)}]{Huertas-Company+2022}
{Huertas-Company}, M., \& {Lanusse}, F. 2022, arXiv e-prints, arXiv:2210.01813.
\newblock \doarXiv{2210.01813}

\bibitem[{{Juneau} {et~al.}(2021){Juneau}, {Olsen}, {Nikutta}, {Jacques}, \&
  {Bailey}}]{Juneau+2021}
{Juneau}, S., {Olsen}, K., {Nikutta}, R., {Jacques}, A., \& {Bailey}, S. 2021,
  Computing in Science and Engineering, 23, 15,
  \dodoi{10.1109/MCSE.2021.3057097}

\bibitem[{{Kauffmann et al.}(2003)}]{Kauffmann+2003}
{Kauffmann et al.}, G. 2003, \mnras, 346, 1055,
  \dodoi{10.1111/j.1365-2966.2003.07154.x}

\bibitem[{{Kewley} {et~al.}(2001){Kewley}, {Dopita}, {Sutherland}, {Heisler},
  \& {Trevena}}]{Kewley+2001}
{Kewley}, L.~J., {Dopita}, M.~A., {Sutherland}, R.~S., {Heisler}, C.~A., \&
  {Trevena}, J. 2001, \apj, 556, 121, \dodoi{10.1086/321545}

\bibitem[{{Kingma} \& {Ba}(2014)}]{Kingma+2014}
{Kingma}, D.~P., \& {Ba}, J. 2014, arXiv e-prints, arXiv:1412.6980.
\newblock \doarXiv{1412.6980}

\bibitem[{Kluyver {et~al.}(2016)Kluyver, Ragan-Kelley, P{\'e}rez, Granger,
  Bussonnier, Frederic, Kelley, Hamrick, Grout, Corlay, Ivanov, Avila, Abdalla,
  \& Willing}]{Kluyver+2016}
Kluyver, T., Ragan-Kelley, B., P{\'e}rez, F., {et~al.} 2016, in Positioning and
  Power in Academic Publishing: Players, Agents and Agendas, ed. F.~Loizides \&
  B.~Schmidt, IOS Press, 87 -- 90

\bibitem[{{McInnes} {et~al.}(2018){McInnes}, {Healy}, \&
  {Melville}}]{McInnes+2018}
{McInnes}, L., {Healy}, J., \& {Melville}, J. 2018, arXiv e-prints,
  arXiv:1802.03426.
\newblock \doarXiv{1802.03426}

\bibitem[{{Nikutta} {et~al.}(2020){Nikutta}, {Fitzpatrick}, {Scott}, \&
  {Weaver}}]{Nikutta+2020}
{Nikutta}, R., {Fitzpatrick}, M., {Scott}, A., \& {Weaver}, B.~A. 2020,
  Astronomy and Computing, 33, 100411, \dodoi{10.1016/j.ascom.2020.100411}

\bibitem[{{Osterbrock} \& {Ferland}(2006)}]{Osterbrock+2006}
{Osterbrock}, D.~E., \& {Ferland}, G.~J. 2006, {Astrophysics of gaseous nebulae
  and active galactic nuclei}

\bibitem[{{Portillo} {et~al.}(2020){Portillo}, {Parejko}, {Vergara}, \&
  {Connolly}}]{Portillo+2020}
{Portillo}, S. K.~N., {Parejko}, J.~K., {Vergara}, J.~R., \& {Connolly}, A.~J.
  2020, \aj, 160, 45, \dodoi{10.3847/1538-3881/ab9644}

\bibitem[{{Richards} {et~al.}(2003){Richards}, {Hall}, {Vanden Berk},
  {Strauss}, {Schneider}, {Weinstein}, {Reichard}, {York}, {Knapp}, {Fan},
  {Ivezi{\'c}}, {Brinkmann}, {Budav{\'a}ri}, {Csabai}, \&
  {Nichol}}]{Gordon+2003}
{Richards}, G.~T., {Hall}, P.~B., {Vanden Berk}, D.~E., {et~al.} 2003, \aj,
  126, 1131, \dodoi{10.1086/377014}

\bibitem[{{Ross} {et~al.}(2015){Ross}, {Hamann}, {Zakamska}, {Richards},
  {Villforth}, {Strauss}, {Greene}, {Alexandroff}, {Brandt}, {Liu}, {Myers},
  {P{\^a}ris}, \& {Schneider}}]{Ross+2015}
{Ross}, N.~P., {Hamann}, F., {Zakamska}, N.~L., {et~al.} 2015, \mnras, 453,
  3932, \dodoi{10.1093/mnras/stv1710}

\bibitem[{{S{\'a}nchez Almeida} {et~al.}(2010){S{\'a}nchez Almeida}, {Aguerri},
  {Mu{\~n}oz-Tu{\~n}{\'o}n}, \& {de Vicente}}]{Almeida+2010}
{S{\'a}nchez Almeida}, J., {Aguerri}, J.~A.~L., {Mu{\~n}oz-Tu{\~n}{\'o}n}, C.,
  \& {de Vicente}, A. 2010, \apj, 714, 487, \dodoi{10.1088/0004-637X/714/1/487}

\bibitem[{{Thomas} {et~al.}(2013){Thomas}, {Steele}, {Maraston}, {Johansson},
  {Beifiori}, {Pforr}, {Str{\"o}mb{\"a}ck}, {Tremonti}, {Wake}, {Bizyaev},
  {Bolton}, {Brewington}, {Brownstein}, {Comparat}, {Kneib}, {Malanushenko},
  {Malanushenko}, {Oravetz}, {Pan}, {Parejko}, {Schneider}, {Shelden},
  {Simmons}, {Snedden}, {Tanaka}, {Weaver}, \& {Yan}}]{Thomas+2013}
{Thomas}, D., {Steele}, O., {Maraston}, C., {et~al.} 2013, \mnras, 431, 1383,
  \dodoi{10.1093/mnras/stt261}

\bibitem[{{Vanden Berk} {et~al.}(2001){Vanden Berk}, {Richards}, {Bauer},
  {Strauss}, {Schneider}, {Heckman}, {York}, {Hall}, {Fan}, {Knapp},
  {Anderson}, {Annis}, {Bahcall}, {Bernardi}, {Briggs}, {Brinkmann}, {Brunner},
  {Burles}, {Carey}, {Castander}, {Connolly}, {Crocker}, {Csabai}, {Doi},
  {Finkbeiner}, {Friedman}, {Frieman}, {Fukugita}, {Gunn}, {Hennessy},
  {Ivezi{\'c}}, {Kent}, {Kunszt}, {Lamb}, {Leger}, {Long}, {Loveday}, {Lupton},
  {Meiksin}, {Merelli}, {Munn}, {Newberg}, {Newcomb}, {Nichol}, {Owen}, {Pier},
  {Pope}, {Rockosi}, {Schlegel}, {Siegmund}, {Smee}, {Snir}, {Stoughton},
  {Stubbs}, {SubbaRao}, {Szalay}, {Szokoly}, {Tremonti}, {Uomoto}, {Waddell},
  {Yanny}, \& {Zheng}}]{Berk+2001}
{Vanden Berk}, D.~E., {Richards}, G.~T., {Bauer}, A., {et~al.} 2001, \aj, 122,
  549, \dodoi{10.1086/321167}

\bibitem[{{York} {et~al.}(2000){York}, {Adelman}, {Anderson}, {Anderson},
  {Annis}, {Bahcall}, {Bakken}, {Barkhouser}, {Bastian}, {Berman}, {Boroski},
  {Bracker}, {Briegel}, {Briggs}, {Brinkmann}, {Brunner}, {Burles}, {Carey},
  {Carr}, {Castander}, {Chen}, {Colestock}, {Connolly}, {Crocker}, {Csabai},
  {Czarapata}, {Davis}, {Doi}, {Dombeck}, {Eisenstein}, {Ellman}, {Elms},
  {Evans}, {Fan}, {Federwitz}, {Fiscelli}, {Friedman}, {Frieman}, {Fukugita},
  {Gillespie}, {Gunn}, {Gurbani}, {de Haas}, {Haldeman}, {Harris}, {Hayes},
  {Heckman}, {Hennessy}, {Hindsley}, {Holm}, {Holmgren}, {Huang}, {Hull},
  {Husby}, {Ichikawa}, {Ichikawa}, {Ivezi{\'c}}, {Kent}, {Kim}, {Kinney},
  {Klaene}, {Kleinman}, {Kleinman}, {Knapp}, {Korienek}, {Kron}, {Kunszt},
  {Lamb}, {Lee}, {Leger}, {Limmongkol}, {Lindenmeyer}, {Long}, {Loomis},
  {Loveday}, {Lucinio}, {Lupton}, {MacKinnon}, {Mannery}, {Mantsch}, {Margon},
  {McGehee}, {McKay}, {Meiksin}, {Merelli}, {Monet}, {Munn}, {Narayanan},
  {Nash}, {Neilsen}, {Neswold}, {Newberg}, {Nichol}, {Nicinski}, {Nonino},
  {Okada}, {Okamura}, {Ostriker}, {Owen}, {Pauls}, {Peoples}, {Peterson},
  {Petravick}, {Pier}, {Pope}, {Pordes}, {Prosapio}, {Rechenmacher}, {Quinn},
  {Richards}, {Richmond}, {Rivetta}, {Rockosi}, {Ruthmansdorfer}, {Sandford},
  {Schlegel}, {Schneider}, {Sekiguchi}, {Sergey}, {Shimasaku}, {Siegmund},
  {Smee}, {Smith}, {Snedden}, {Stone}, {Stoughton}, {Strauss}, {Stubbs},
  {SubbaRao}, {Szalay}, {Szapudi}, {Szokoly}, {Thakar}, {Tremonti}, {Tucker},
  {Uomoto}, {Vanden Berk}, {Vogeley}, {Waddell}, {Wang}, {Watanabe},
  {Weinberg}, {Yanny}, {Yasuda}, \& {SDSS Collaboration}}]{York+2000}
{York}, D.~G., {Adelman}, J., {Anderson}, Jr., J.~E., {et~al.} 2000, \aj, 120,
  1579, \dodoi{10.1086/301513}

\end{thebibliography}

\end{document}